\documentclass[a4paper,10pt,twocolumn,twoside,journal]{article}

\usepackage{geometry}
\advance\oddsidemargin by -1in
\advance\evensidemargin by -1in
\advance\headsep by 2.8mm
\usepackage{float}
\usepackage{cite}
\usepackage{authblk}
\usepackage{amsmath,amssymb,amsfonts}
\usepackage{amsthm}
\usepackage{multirow}
\usepackage{booktabs}
\usepackage{hyperref}
\usepackage{algorithm}
\usepackage{algorithmicx}
\usepackage{algpseudocode}
\usepackage{graphicx}
\usepackage{textcomp}
\usepackage{array}
\usepackage{caption}
\usepackage{subcaption}
\usepackage{caption,setspace}
\usepackage{wrapfig}
\def\BibTeX{{\rm B\kern-.05em{\sc i\kern-.025em b}\kern-.08em
    T\kern-.1667em\lower.7ex\hbox{E}\kern-.125emX}}
    
\usepackage{xcolor}
\usepackage{comment}
\usepackage{braket}   
\usepackage{pgfplots}
\usepackage{pgfplotstable}
\pgfplotsset{compat=1.7}

\usepackage{tikz}

\usetikzlibrary{arrows.meta}
\usetikzlibrary{quantikz}
\usepackage{adjustbox}

\usepackage[normalem]{ulem}

\theoremstyle{remark}
\newtheorem{theo}{Theorem}

\theoremstyle{definition}

\title{Compiler Design for Distributed Quantum Computing}

\author[1]{Davide~Ferrari}
\author[2,3]{Angela~Sara~Cacciapuoti}
\author[1]{Michele~Amoretti}
\author[2,3]{Marcello~Caleffi}
\affil[1]{\small \textit{Quantum Software}, Department of Engineering and Architecture (DIA), University of Parma, Parma, 43124 Italy (e-mail: \href{mailto:michele.amoretti@unipr.it}{michele.amoretti@unipr.it}, \href{mailto:davide.ferrari1@unipr.it}{davide.ferrari1@unipr.it}). Web: \href{http://www.qis.unipr.it/quantumsoftware.html}{http://www.qis.unipr.it/quantumsoftware.html}}
\affil[2]{\small \textit{FLY: Future Communications Laboratory}, Department of Electrical Engineering and Information Technology (DIETI), University of Naples Federico II, Naples, 80125 Italy (e-mail: \href{mailto:marcello.caleffi@unina.it}{marcello.caleffi@unina.it}, \href{mailto:angelasara.cacciapuoti@unina.it}{angelasara.cacciapuoti@unina.it}). Web: \href{http://www.quantuminternet.it}{www.quantuminternet.it}.}
\affil[3]{\small \textit{Laboratorio Nazionale di Comunicazioni Multimediali}, National Inter-University Consortium for Telecommunications (CNIT), Naples, 80126 Italy.}

\date{}

\begin{document}

\maketitle

\begin{abstract}
In distributed quantum computing architectures, with the network and communications functionalities provided by the Quantum Internet, remote quantum processing units (QPUs) can communicate and cooperate for executing computational tasks that single NISQ devices cannot handle by themselves. To this aim, distributed quantum computing requires a new generation of quantum compilers, for mapping any quantum algorithm to any distributed quantum computing architecture. With this perspective, in this paper, we first discuss the main challenges arising with compiler design for distributed quantum computing. Then, we analytically derive an upper bound of the overhead induced by quantum compilation for distributed quantum computing. The derived bound accounts for the overhead induced by the underlying computing architecture \textit{as well as} the additional overhead induced by the sub-optimal quantum compiler -- expressly designed through the paper to achieve three key features, namely, \textit{general-purpose}, \textit{efficient} and \textit{effective}. Finally, we validate the analytical results and we confirm the validity of the compiler design through an extensive performance analysis.
\end{abstract}


\maketitle

\section{Introduction}
\label{sec:1}

Current quantum computers are commonly defined as \textit{noisy intermediate-scale quantum (NISQ)} devices, being characterized by few dozens of quantum bits (qubits) with non-uniform quality and highly constrained physical connectivity.

Hence, the growing demand for large-scale quantum computers is motivating research on distributed quantum computing architectures \cite{CalCacBia-18, CalChaCuo-20}, and experimental efforts have demonstrated some of the building blocks for such a design \cite{VanDev-16}. Indeed, with the network and communications functionalities provided by the \textit{Quantum Internet} \cite{PirBra-16,Gib-16,DurLamHeu-17,Sim-17,Wehner2018,Zomo2018,CalCacBia-18,CalChaCuo-20,GyoImr-20,GyoImr-20-2}, remote quantum processing units (QPUs) can communicate and cooperate -- through the distributed computing paradigm as a \textit{virtual quantum processor} with a number of qubits that scales linearly with the number of remote QPUs \cite{CuoCalCac-20} -- for executing computational tasks that each NISQ device cannot handle by itself. 

As overviewed in recent literature such as \cite{VanDev-16,CuoCalCac-20}, several challenges arise with the design of a distributed quantum computing architecture. In the following, we focus on the problem of designing a \textit{quantum algorithm compiler} for distributed quantum compilation.

Compiling a quantum algorithm means translating a hardware-agnostic description of the algorithm -- i.e., the quantum circuit\footnote{See Section~\ref{sec:2} for a proper introduction to quantum circuits, circuit compilation and circuit depth.} -- into a functionally-equivalent one that takes into account the physical constraints of the underlying computing architecture -- i.e., the \textit{compiled} quantum circuit. When it comes to distributed computing architectures, two are the main issues arising with the compiler design.

First, a fundamental question arises with distributed computation: \textit{at what price}? Indeed, distributed computation requires the different processors being able to communicate each others for coordinating and data exchanging, and these tasks introduce an overhead that strongly depend on the particulars of the distributed computing architecture. For instance, the induced overhead becomes more severe as the connectivity between the QPUs shrinks or as the number of qubits stored at the QPUs decreases. Hence, from a compiling perspective, it is crucial to estimate the overhead effects onto the compiled quantum circuit, effects that are generally measured in terms of \textit{depth} of the compiled circuit with respect to the depth of the original one.

Furthermore, compiling a quantum circuit is a very challenging task even for a single-processor architecture, being such a task a NP-complete problem \cite{BotKisMar-18}. Hence, optimal circuit compiling for distributed quantum architectures can be achieved only for very small circuit instances. Conversely, the compilation of medium-to-large circuits of practical value induces an additional overhead -- whose severity depends on the suboptimality of the quantum compiler -- that further increases the depth of the compiled circuit.

With this in mind, in this paper we analytically derive an upper bound of the overhead induced by quantum circuit compilation for distributed quantum computing:
\begin{itemize}
    \item by considering the overhead induced by the worst-case scenario for a distributed quantum computing architecture, namely a scenario characterized by i) the lowest possible number of qubits at each QPU, and ii) the \textit{poorest} connectivity among the QPUs,
    \item and by considering the additional overhead induced by a sub-optimal quantum compiler.
\end{itemize}

Clearly, with reference to the last point, the additional overhead strongly depends on the particulars of the quantum compiler. To this aim, through the manuscript we design a quantum compiler with three key features:
\begin{itemize}
    \item  \textit{general-purpose}, namely, requiring no particular assumptions on the quantum circuits to be compiled;
    \item \textit{efficient}, namely, exhibiting a polynomial-time computational complexity so that it can successfully compile medium-to-large circuits of practical value;
    \item \textit{effective}, being the total circuit depth overhead induced by the quantum circuit compilation always upper-bounded by a factor that grows linearly with the number of logical qubits of the original quantum circuit.
\end{itemize}

The rest of the paper is organized as follows. In Section~\ref{sec:2} we review some preliminaries about quantum circuits and quantum compilers. Then, in Section~\ref{sec:3} we detail the problem of circuit compilation for distributed quantum computing, discussing the challenges that arise with the compiler design and the relevant literature. These basics are crucial for understanding the compiler design as well as the analytical derivation of the overhead bound given in Section~\ref{sec:4}. Then, Section~\ref{sec:5} presents the performance analysis for the proposed compiler design. In particular, we present the implementation of a compiler that is able to cope with the worst-case scenario for a distributed quantum computing architecture; we validate the compiling overhead upper bound; we illustrate experimental results regarding the compilation of several quantum circuits, with our compiler compared to a state-of-the-art solution. Finally, Section~\ref{sec:6} concludes the paper.

\section{Background}
\label{sec:2}

We refer the reader to \cite{RiePol-11} for an introduction to the conceptual and notation differences separating quantum computing from conventional computing, and to \cite{NieChu-10} for an in-depth treatise of the subject.

In this work, we consider QPUs that support the \textit{quantum circuit model} \cite{Deutsch1985}, which is the most popular and developed model for quantum computation.
A quantum circuit is a model of a quantum algorithm, where quantum operators are described as \emph{quantum gates}. A quantum circuit is still a logical abstraction, not to be confused with its realization on an actual quantum hardware device.
Hence, in the following, the abstract qubits subjected to quantum gates as specified by the quantum circuit are called \textit{logical qubits} to distinguish them from the \textit{physical qubits} embedded within a quantum processor.

\begin{figure}[t!]
    \begin{adjustbox}{width=\columnwidth}
        \includegraphics[]{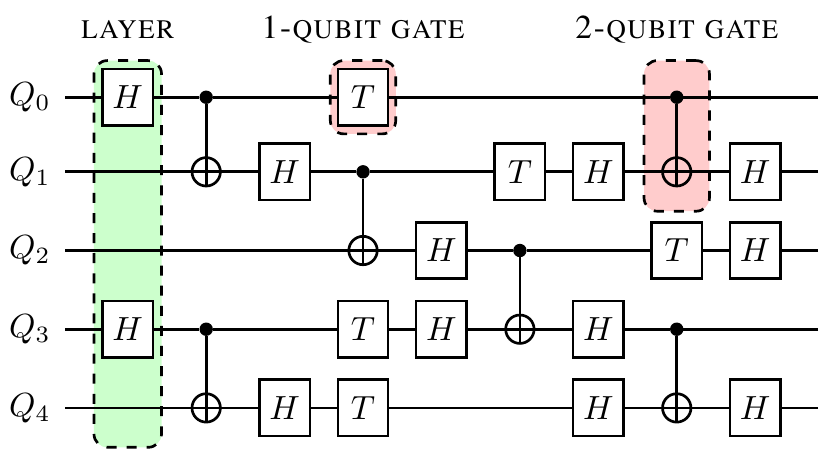}
	\end{adjustbox}
    \caption{Example of a 5-qubit quantum circuit from \cite{BoiIsaSme-18}, with each horizontal line representing the time-evolution of the state of a single logical qubit.}
    \label{Fig:01}
	\hrulefill
\end{figure}

Figure~\ref{Fig:01} shows a simple quantum circuit, where each horizontal line represents the time-evolution of the state of a single logical qubit, with time flowing from left to right, dictating the order of execution of the different gates. More specifically, gates affecting the same qubit must be executed sequentially, and this agrees with the intuition. Conversely, gates acting on different qubits can be performed simultaneously as long as the ``ordering'' arising from gates affecting multiple qubits is respected.
This concept underlies the notion of \textit{layer}, i.e., the set of gates that can be performed simultaneously on a disjoint set of qubits. The number of layers in a quantum circuit is denoted as \textit{circuit depth}.
As an example, the quantum circuit given in Figure~\ref{Fig:01} is composed of $9$ layers and hence its depth is equal to $9$. The number of gates within the circuit is denoted as \textit{circuit size}.

\subsection{Quantum Compilation}
\label{sec:2.2}

Given a quantum algorithm, there exist several equivalent quantum circuits modeling the same computation with a different arrangement or different ordering of gates.

Circuits with fewer gates -- i.e., with lower \textit{size} -- may be preferred to reduce the circuit complexity. However, the execution time of the circuit -- rather than its size -- is generally considered the key factor to be optimized \cite{KanTemCor-19,GyoImr-20-3}. The rationale is to keep the execution time of the quantum circuit within the coherence time of the underlying quantum hardware architecture \cite{NieChu-10,CacCalVan-20}. By oversimplifying, the execution time increases with the number of layers. Therefore, it is crucial to build -- for a given quantum algorithm -- a quantum circuit characterized by the lowest possible depth. However, two issues arise as a consequence of the quantum processor characteristics.

First, even if there exists an uncountable number of quantum logic gates, the set of gates that can be executed on a certain quantum processor can be limited, as a consequence of the constraints imposed by the underlying qubit technology \cite{VanDev-16}. In this case, any gate outside this \textit{reduced set} must be obtained with a proper combination of the allowed gates through a process known as \textit{gate synthesis}.

\begin{figure}
    \begin{adjustbox}{width=\columnwidth}
        \includegraphics[]{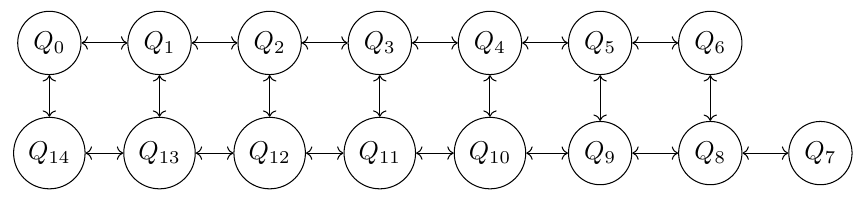}
	\end{adjustbox}
    \caption{Coupling map of the IBM Melbourne quantum processor \cite{IBMQ-backends}. The fifteen physical qubits are represented by circles. The arrows denote the possibility to realize a two-qubit \texttt{CNOT} gate between the connected qubits, with the arrow pointing toward the target qubit. As an example, a \texttt{CNOT} between qubits $Q_1$ (control) and $Q_0$ (target) can be directly executed by the quantum processor, whereas a  \texttt{CNOT} between qubits $Q_2$ and $Q_0$ cannot.}
    \label{Fig:02}
	\hrulefill
\end{figure}

Furthermore, regardless of the underlying qubit technology, any quantum processor exhibits physical constraints -- arising as a consequence of the noise and the physical-space limitations -- on the possible interactions between the different physical qubits. For example, \texttt{CNOT} gates cannot be applied to any physical qubit pair, but they are instead restricted to certain pairs, as shown in Figure~\ref{Fig:02} with the \textit{coupling map} of an IBM quantum processor.

\begin{figure*}[ht]
	\centering
        \begin{minipage}[c]{.29\linewidth}
		\centering
        \begin{adjustbox}{width=1\linewidth}
            \includegraphics[]{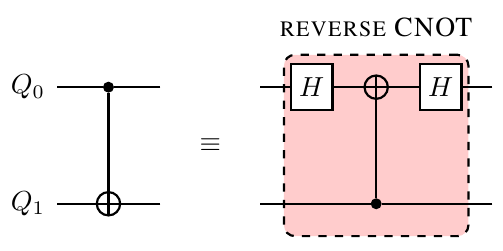}
        \end{adjustbox}
	    \subcaption{Reversing \texttt{CNOT} \cite{GarCha-11}. A \texttt{CNOT} between $Q_0$ (control) and $Q_1$ (target) can be executed with the coupling map given in Figure~\ref{Fig:02} by performing a \texttt{CNOT} between $Q_1$ (control) and $Q_0$ (target) sandwiched between two $H$ gates. We note that IBM Melbourne processor (shown in Figure~\ref{Fig:02}) natively supports \texttt{CNOT}s in both directions between neighbor qubits.}
        \label{Fig:03-a}
    \end{minipage}
    \hfil
    \begin{minipage}[c]{.69\linewidth}
		\centering
		\begin{adjustbox}{width=1\linewidth}
            \includegraphics[]{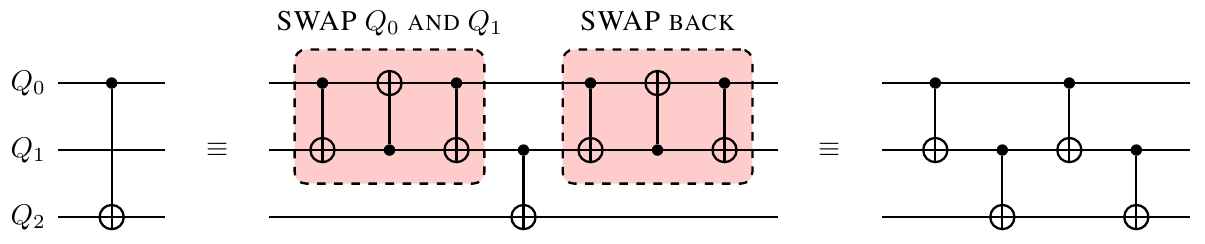}
        \end{adjustbox}
	    \subcaption{ 
	    A \texttt{CNOT} between qubits $Q_0$ (control) and $Q_2$ (target) can be executed 
	    through either: i) \textit{quantum state transfer}, by first swapping qubits $Q_0$ and $Q_1$ so that $Q_0$ and $Q_2$ become adjacent qubits in the coupling maps, then by performing a \texttt{CNOT} between $Q_0$ and $Q_2$, and finally by swapping again qubits $Q_0$ and $Q_1$ so that they recover their initial position, or ii) \textit{ancilla qubit}, by performing four \texttt{CNOT} operations between neighbour qubits with the help of the intermediate qubit $Q_1$.} 
		\label{Fig:03-b}
	\end{minipage}
    \caption{Example of equivalent quantum circuits generated during the quantum compilation for mapping an arbitrary \texttt{CNOT} into a sequence of \texttt{CNOT}s that can be directly executed by a given quantum processor.}
    \label{Fig:03}
    \hrulefill
\end{figure*}

From the above, it becomes clear that the execution of a quantum algorithm on a certain quantum processor requires that: i) each logical qubit of the quantum circuit is mapped
\footnote{Indeed, NISQ technology may require a logical qubit to be mapped onto several physical qubits to implement proper fault-tolerant techniques \cite{CorKanJav-20}. Nevertheless, in the following we assume a one-to-one mapping for the sake of clarity, without any loss of generality.} onto a physical qubit of the quantum processor, and ii) each \texttt{CNOT} operation between non-adjacent (within the coupling map) physical qubits is mapped into a sequence of \texttt{CNOT} operations between adjacent physical qubits, as shown\footnote{With the \textit{state transfer} strategy based on \texttt{SWAP}s usually preferred over the \textit{ancilla} strategy \cite{ZulPalWil-19,Li2019,Qiskit}.} in Figure~\ref{Fig:03}. 

This process, known as \textit{quantum compilation}, must be optimized so that the depth of the \textit{compiled circuit} -- i.e., the equivalent quantum circuit satisfying all the constrains imposed by the quantum processor -- is minimized \cite{FerAmo-18,CinSubSor-18,ZulPalWil-19,Li2019}.

\section{Compilers for Distributed Quantum Computing}
\label{sec:3}

As highlighted in Section~\ref{sec:1}, the demand for large-scale quantum computers is motivating research on distributed quantum computing architectures, where multiple small-scale quantum processors interact and cooperate through the Quantum Internet for solving challenging computational tasks. As a consequence, a new generation of quantum compilers is needed, for mapping any quantum algorithm to any distributed quantum computing architecture.

\begin{figure*}
	\centering
        \begin{minipage}[c]{.49\linewidth}
		\centering
        \begin{adjustbox}{width=\columnwidth}
            \includegraphics[]{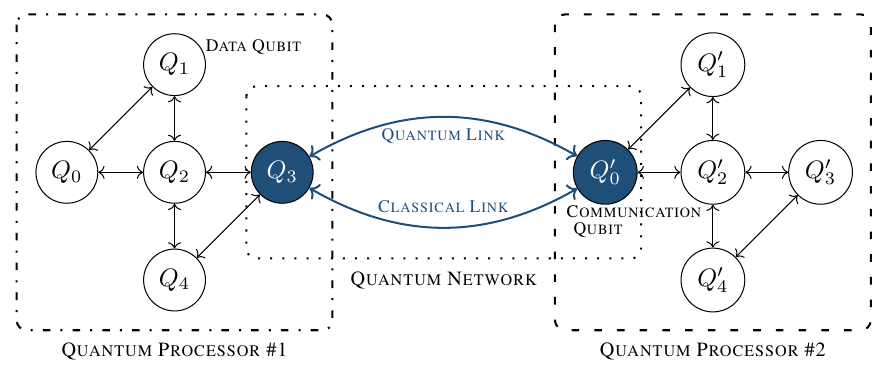}
    	\end{adjustbox}
	    \subcaption{Two IBM Yorktown quantum processors are interconnected with a quantum link and a classical link. The classical link is used to transmit classical information, whereas the quantum link is needed to distribute \textit{Bell states} -- that is, maximally-entangled two-qubit states -- between remote processors to execute remote operations. Indeed, at least one physical qubit at each processor must be reserved for storing the Bell state, as discussed in Figure~\ref{Fig:04-b}. This kind of qubits -- dark-blue-colored in the figure -- are called \textit{communication qubits} \cite{CalChaCuo-20,KozWehVan-20} to distinguish them from the remaining physical qubits -- white-colored in the figure --  devoted to computing and referred to as \textit{data qubits}.}
        \label{Fig:04-a}
    \end{minipage}
    \hfil
    \begin{minipage}[c]{.49\linewidth}
		\centering
		\begin{adjustbox}{width=1\linewidth}
		  \includegraphics[]{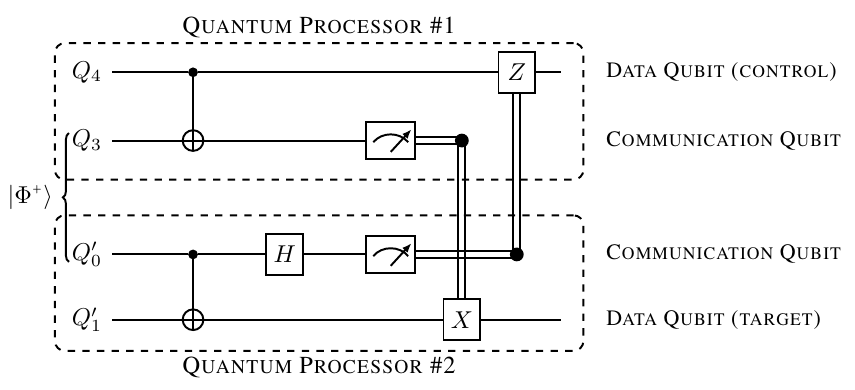}
        \end{adjustbox}
	    \subcaption{Remote \texttt{CNOT}. To perform a \texttt{CNOT} between remote physical qubits stored at different processors -- say data qubits $Q_4$ and $Q'_1$ in Figure~\ref{Fig:04-a} -- a Bell state such as $\ket{\Phi^{\texttt{+}}}$  must be distributed through the quantum link so that each pair member is stored within a \textit{communication qubit} at each processor. Once the Bell state is available, the remote \texttt{CNOT} is obtained with a local \texttt{CNOT} between the \textit{data} and the \textit{communication qubit} at each processor, followed by a conditional gate on the data qubit depending on the measurement of the remote communication qubit. The double line denotes the transmission of one bit of classical information -- i.e., the measurement output -- between the remote processors.}
		\label{Fig:04-b}
	\end{minipage}
    \caption{Toy-model for distributed quantum computing, with two quantum processors interconnected through a quantum network. Figure~\ref{Fig:04-a} shows the network topology along with the processors coupling maps, whereas Figure~\ref{Fig:04-b} provides the quantum circuit detailing the classical (2 bits) and the quantum (the Bell state) resources needed to execute a remote operation.}
    \label{Fig:04}
    \hrulefill
\end{figure*}

Let us consider a toy model for distributed quantum computing, in which a generic quantum algorithm must be executed on two quantum processors interconnected by a quantum link, as shown in Figure~\ref{Fig:04}.

\subsection{Challenges}
\label{sec:3.1}
Several challenges arise with the design of a quantum compiler for mapping an arbitrary quantum circuit into a distributed quantum computing architecture, as discussed in the following.

\subsubsection*{Data Qubits vs Communication qubits}
Similarly to classical distributed computing, a key requirement for distributed quantum computing is the possibility to perform \textit{remote operations}, namely operations between qubits stored at different processors. But, differently from the classical domain, quantum mechanics does not allow an unknown qubit to be copied or even simply read or measured in any way, without causing an irreversible loss of the quantum information stored within the qubit \cite{NieChu-10,RiePol-11}.

Thankfully, \textit{entanglement} provides an invaluable tool for implementing remote operations without violating quantum mechanics \cite{CuoCalCac-20}. Entanglement is a property of two (or more, in case of \textit{multipartite} entanglement) quantum particles that exist in a special type of superposition state, such that any action on a particle affects instantaneously the other particle as well. This sort of quantum correlation, with no counterpart in the classical world, holds even when the particles are far away from each other. For an in-depth discussion about entanglement from an information engineering point of view, we refer the reader to \cite{CacCalVan-20}.

By exploiting the availability of a \textit{Bell state} -- that is a state of two maximally-entangled qubits -- shared between the two remote processors, it is possible to perform a remote \texttt{CNOT} through a sequence of local \texttt{CNOT}s and single-qubit operations/measurements as shown in Figure~\ref{Fig:04-b}. 

To distribute Bell states between different quantum processors, at least one qubit at each processor -- referred to as \textit{communication qubit} \cite{KozWehVan-20} to distinguish it from the remaining \textit{data qubits} devoted to processing -- must be reserved for remote inter-processor operations. Hence, a crucial trade-off between communication and data qubits arises. Specifically, for each remote \texttt{CNOT}, a Bell state is consumed (Figure~\ref{Fig:04-b}) and a new Bell state must be distributed between the remote processors through the quantum link before another remote \texttt{CNOT} can be executed. Hence, the more communication qubits are available within a processor, the more remote \texttt{CNOT}s can be executed in parallel, reducing the overhead induced by the distributed computation. But the more communication qubits are available for inter-processor communication, the less valuable resources -- i.e., data qubits -- are available for computing.

It is unlikely that a data qubit could be dynamically turned into a communication qubit during the compilation, given the dedicated hardware -- such as a matter-flying qubit interface \cite{CacCalVan-20} -- required for entanglement distribution. Conversely, it is reasonable to envision that the distributed quantum compiler could easily reserve -- when multiple communication qubits are available at the same processor -- a subset of the communication qubits for computing. This optimization task represents an interesting yet unaddressed open problem.

\subsubsection*{Dynamic Connectivity}

\begin{figure}[t!]
	\centering
    \begin{adjustbox}{width=1\columnwidth}
        \includegraphics[]{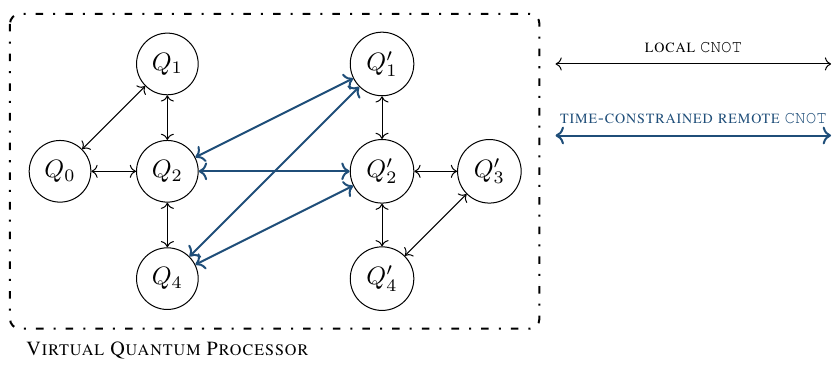}
    \end{adjustbox}
    \caption{Dynamic coupling map for the network topology shown in Figure~\ref{Fig:04-a}. The two 5-qubit quantum processors constitute a 8-qubit \textit{virtual quantum processor} with qubits inter-connected through both local and remote \texttt{CNOT}s. While the local \texttt{CNOT}s can be concurrently executed -- i.e., they are \textit{unconstrained} -- the parallel execution of multiple remote \texttt{CNOT}s is \textit{constrained} to the availability of multiple Bell states, with one Bell state for each concurrent remote \texttt{CNOT}. Given that only one communication qubit is available at each processor in Figure~\ref{Fig:04-a}, out of four remote \texttt{CNOT}s only one can be executed at any time.}
    \label{Fig:05}
    \hrulefill
\end{figure}

As mentioned in Section~\ref{sec:2.2}, with single-processor quantum computing all the constraints on the possible interactions between different qubits -- arising from the underlying physical computing architecture -- can be effectively represented with a coupling map. Formally, a coupling map is a visual representation of the directed graph $G$:
\begin{equation}
    \label{eq:3.1}
    G=(\mathcal{V},\mathcal{E})
\end{equation}
where $\mathcal{V} = \{v_i \}$ denotes the set of vertices representing the qubits and $\mathcal{E} = \{e_{i,j}\}_{v_i,v_j \in V}$ denotes the set of directed edges representing the possibility to perform\footnote{In the following, for the sake of presentation, we consider the simplest binary case, i.e., either the operation can or cannot be executed. But the discussions, as well as the results derived in the following, continue to hold when a weight -- usually representing the gate fidelity -- is mapped on the edge.} 
a \texttt{CNOT} with $v_i$ and $v_j$ acting as control and target qubit, respectively. 

But when it comes to distributed quantum computing, a new kind of constraints arises as a consequence of the underlying physical network topology.

More in detail, similarly to single-processor quantum compiling, the remote operations are restricted to certain fixed pairs. Specifically, they are restricted to pairs composed by data qubits directly connected to a communication qubit within the processor coupling map. For instance, with reference to the network topology shown in Figure~\ref{Fig:04-a}, a remote \texttt{CNOT} between data qubits $Q_2$ and $Q'_1$ can be directly mapped onto the circuit given in Figure~\ref{Fig:04-b}. Conversely, a remote \texttt{CNOT} between data qubits $Q_2$ and $Q'_4$ cannot be directly executed between the pair, but it requires to distribute the operation through the neighbor qubits as shown in Figure~\ref{Fig:03-b}.

However, differently from single-processor compiling, the remote operations are subjects to two types of \textit{temporal constraints}.
\begin{itemize}
    \item \textit{Simultaneity Limitations}. As previously discussed, each remote \texttt{CNOT} relies on the availability of a Bell state stored within a communication qubit. Hence, even if remote \texttt{CNOT}s can -- in principle -- be executed between different remote pairs, the number of remote \texttt{CNOT}s that can be executed simultaneously between two processors is limited by the number of communication qubits jointly available at each processor. With reference to Figure~\ref{Fig:05}, out of four possible remote \texttt{CNOT}s (denoted with blue arrows), only one can be executed at any time.
    \item \textit{Consecutiviness Limitations}. Each remote \texttt{CNOT} consumes a Bell state as a consequence of the measurement operations on the communication qubits \cite{CacCalVan-20}. Accordingly, a new Bell state must be generated and distributed through the quantum link to the communication qubits, before a subsequent remote \texttt{CNOT} could be executed. And even if the Bell state distribution can start right after the measurements, it is reasonable to assume -- given the several order of magnitudes separating intra-processor qubit distance from inter-processor one -- that the time needed to entangle the communication qubits significantly exceeds the time required for local \texttt{CNOT}s\footnote{In the order of hundreds of nanoseconds for the IBM Yorktown quantum processor \cite{LinMasRoe-17}}. Accordingly, we have two major issues. First, the ``clock'' of the remote operations will be significantly lower than the ``clock'' of the local operations and, hence, it becomes fundamental to minimize the number of remote -- rather than the number of local -- operations to preserve the quantum information integrity from decoherence (Section~\ref{sec:2.2}). Furthermore, there may be periods of time -- following the execution of a remote operation up to the successful distribution of a new Bell state -- during which the quantum processors are \textit{disconnected} and only local operations are possible.
\end{itemize}

These additional constraints must be properly modeled within the coupling map, so that the distributed quantum compiler can optimize the quantum circuit by accounting for the temporal dynamics arising with the distributed architecture. And this represents an open problem.

\begin{figure*}[ht!]
    \begin{minipage}[c]{1\linewidth}
		\centering
        \begin{adjustbox}{width=1\linewidth}
            \includegraphics[]{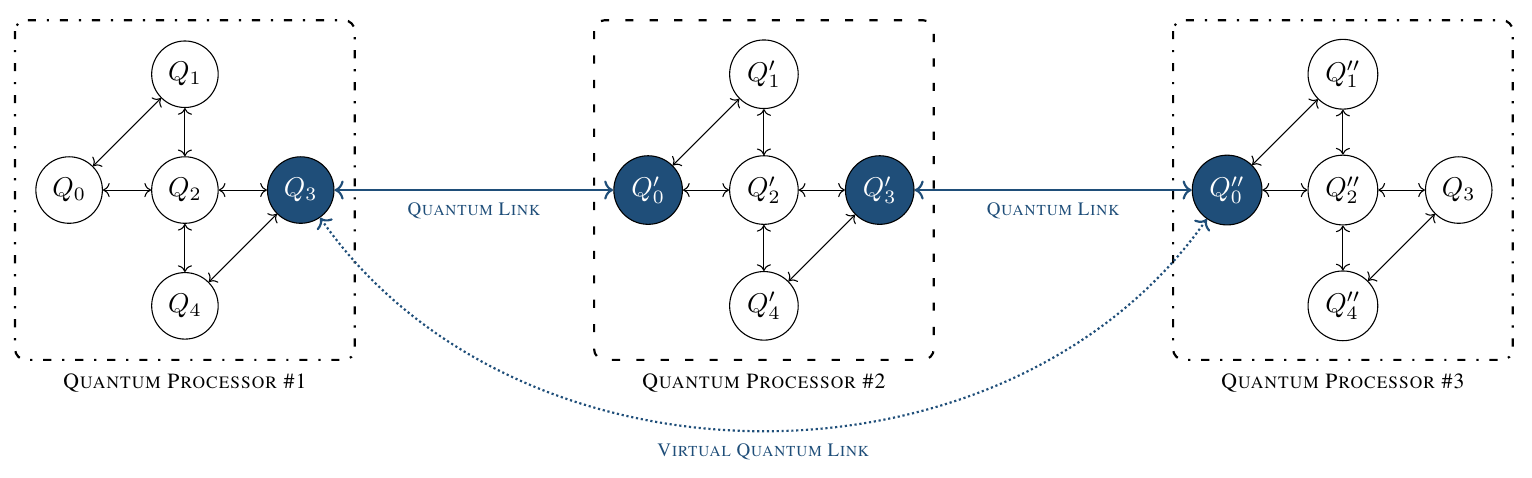}
    	\end{adjustbox}
        \subcaption{By swapping the entanglement at the intermediate nodes -- namely, quantum processor \#2 -- it is possible to distribute a Bell state between remote processors -- namely, processors \#1 and \#3 -- even if they are not adjacent, i.e., they are not directly connected through a quantum link. Hence, entanglement swapping enhances the network connectivity through \textit{virtual quantum links}.\\
        }
        \label{Fig:06-a}
    \end{minipage}
    
    \begin{minipage}[c]{.35\linewidth}
		\centering
		\begin{adjustbox}{width=.8\linewidth}
		  \includegraphics[]{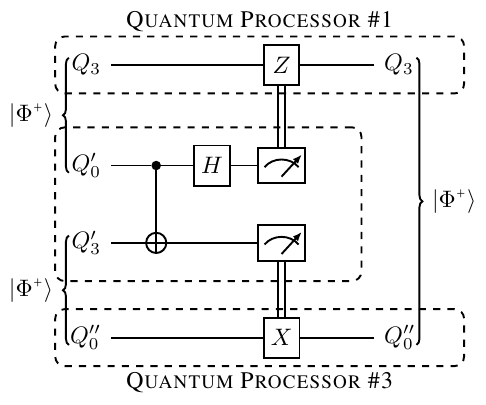}
        \end{adjustbox}
        \subcaption{Entanglement swapping. A Bell state can be distributed between remote processors by swapping the entanglement at an intermediate node through local processing and classical communication.}
        \label{Fig:06-b}
    \end{minipage}
    \hfil
    \begin{minipage}[c]{.63\linewidth}
		\centering
		\begin{adjustbox}{width=1\linewidth}
            \includegraphics[]{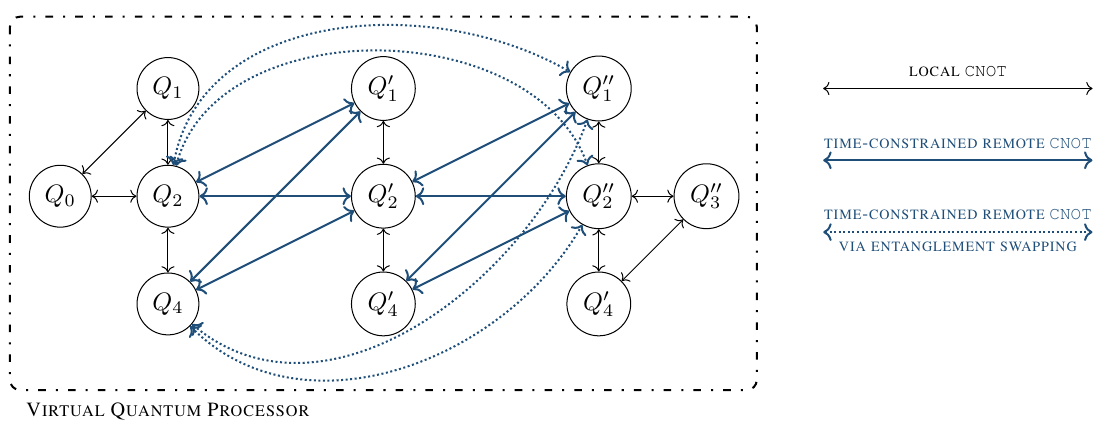}
        \end{adjustbox}
	    \subcaption{Dynamic coupling map for the network topology shown in Figure~\ref{Fig:06-a}. The solid blue lines denote remote \texttt{CNOT}s between adjacent processors, whereas the dotted blue lines denote remote \texttt{CNOT}s between distant processors achievable via entanglement swapping.} 
		\label{Fig:06-c}
	\end{minipage}
    \caption{Augmented connectivity. Entanglement swapping increases the connectivity between physical qubits, with a number of possible remote \texttt{CNOT}s that scales at least linearly with the number of processors.}
    \label{Fig:06}
    \hrulefill
\end{figure*}

\subsubsection*{Augmented Connectivity}

As shown in Figure~\ref{Fig:03-b}, single-processor quantum computing must resort to either \textit{state transfer} (swapping) or \textit{ancilla} strategy to implement a \texttt{CNOT} between non-adjacent (within the coupling map) physical qubits. The rationale for this lays in the impossibility to have direct interactions between distant qubits. And the further the qubits are within the coupling map, the longer the sequence of additional \texttt{CNOT}s is required, regardless of the adopted strategy.

Conversely, distributed quantum computing can exploit a strategy -- called \textit{entanglement swapping} \cite{VanDev-16} and summarized in Fig~\ref{Fig:06} -- to implement a remote \texttt{CNOT} between qubits stored at remote processors, even if the processors are not directly connected through a quantum link.

In a nutshell, to distribute a Bell state between remote processors -- say quantum processor \#1 and \#3 in Figure~\ref{Fig:06-a} -- two Bell states must be first distributed through the quantum links so that one Bell state is shared between the first processor and an intermediate node and another Bell state is shared by the same intermediate node and the second processor.  Then, by performing a Bell state measurement (consisting of a \texttt{H} and a \texttt{CNOT} gate, followed by a joint measurement) on the communication qubits at the intermediate node -- i.e., qubits $Q'_0$ and $Q'_3$ in Figure~\ref{Fig:06-b} -- a Bell state is obtained at the remote communication qubits -- i.e., qubits $Q''_0$ and $Q_3$ in Figure~\ref{Fig:06-a} -- by applying some local processing at the remote nodes depending on the (classical) output of the Bell state measurement.

From the above, it becomes clear that entanglement swapping significantly increases the connectivity within the virtual quantum processor. As an instance, qubit $Q_4$ in Figure~\ref{Fig:06-a} can interact with just two qubits within the same processor via local \texttt{CNOT}s and two qubits within the neighbor processor via remote \texttt{CNOT}s. However, it can interact with two more qubits -- i.e.,  $Q''_1$ and  $Q''_2$ -- via entanglement swapping. And the higher is the number of available quantum processors, the higher is the number of possible interactions. Indeed, the number of additional interactions via entanglement swapping scales linearly with the number of available processors when only two communication qubits are available at each intermediate processor. If this constraint is relaxed, the number of additional interactions via entanglement swapping scales more than linearly.

However, it must be acknowledged that the augmented connectivity provided by entanglement swapping does not come for free. Indeed, entanglement swapping consumes the Bell states stored within the communication qubits at the intermediate processors. And the higher the number of intermediate processors, the higher the number of consumed Bell states.

Hence, a trade-off between ``augmented connectivity'' and ``EPR cost'' arises with entanglement swapping, and a distributed quantum compiler must carefully account for these pros and cons.

\subsection{Related Work}
\label{sec:3.3}
Most quantum computer proposals are based on variations of the nearest-neighbor, two-qubit, and concurrent execution (NTC) architecture \cite{VanItoh-2005}. Depending on the layout of qubits, there are three NTC architectures: 1D, 2D and 3D. The 1D model, called Linear Nearest Neighbor (LNN) \cite{Fowler2004}, consists of qubits located in a single line. In this model, only two neighboring qubits can interact. This is the most challenging scenario. The effects of the LNN model on performance have been investigated for many relevant use cases, such as the quantum Fourier transform \cite{Takahashi2007, VanMeter2004}, Shor's algorithm \cite{Kutin2007, Van08}, and adders \cite{ChoiVan-2010}.

Beals \textit{et al.} \cite{Beals2013} provided algorithms for efficiently moving and addressing quantum memory in parallel. These imply that the standard circuit model can be simulated with low overhead by a more realistic model of a distributed quantum computer. The authors show that for an LNN $N$-qubit architecture, $O(N)$ time steps are necessary for performing $N/2$ two-qubit gates in parallel. However, it is worthwhile to note that the developed analysis does not consider any additional overhead induced by the compilation task and the derived Big-O bound relies on linear constant that is in the "many, many thousands" \cite{CorLeiRiv-01}. Conversely, in the following we develop an analysis that explicitly considers the additional overhead induced by the compilation task, as discussed in Section~\ref{sec:1}.

Zomorodi-Moghadam \textit{et al.} \cite{Zomo2018} proposed a general approach, based on the Kernighan-Lin algorithm for graph partitioning, to optimize the number of teleportations for a DQC consisting of two spatially separated and long-distance quantum subsystems. The same authors proposed also an approach based on dynamic programming \cite{Dava2020}.

Andr\'es-Mart\'inez and Heunen \cite{AndresMartinez2019} proposed an approach that may distribute circuits across any number of quantum devices. The main idea is to turn the quantum circuit into a hypergraph, then find a partitioning that minimizes the number of cuts, as each cut corresponds to a Bell state shared across two QPUs by means of communication qubits. The partitioning problem is addressed by means of the KaHyPar solver \cite{Akhremtsev2017}. 
The proposed solution has some drawbacks, in particular that there is no way to define the number of communication qubits of each QPU. In the software implementation of the algorithm, the number of available communication qubits is unlimited and cannot be constrained.

\section{Compiler Design and Overhead Bounds}
\label{sec:4}

As discussed in Section~\ref{sec:3}, several additional constraints arise with the shift from single-processor to distributed quantum compiling. And given that single-processor quantum compiling has been already proved to be NP-complete \cite{BotKisMar-18}, it is reasonable to expect that optimal distributed quantum compiling is an even harder challenging task.

For this reason, in the following, we take a completely different approach. Specifically, we aim at designing a \textit{general-purpose}, \textit{efficient} and \textit{effective} compiler for distributed quantum computing.

\textit{General-purpose} because our compiler does not require any particular assumption on the quantum circuit to be compiled.

\textit{Efficient} because -- as proved in Section~\ref{sec:5.1} -- our compiler is computationally efficient, exhibiting a polynomial time complexity that grows polynomially with the number of logical qubits and linearly with the depth of the quantum circuit to be compiled.

\textit{Effective} because -- as proved in Section~\ref{sec:4.2} -- our compiler assures a polynomial worst-case overhead, in terms of both: i) depth of the compiled quantum circuit and ii) number of calls to the costliest and most challenging task, i.e., the link entanglement generation.

\subsection{System Model}
\label{sec:4.1}

\begin{figure*}
	\centering
	\begin{adjustbox}{width=1\linewidth}
        \includegraphics[]{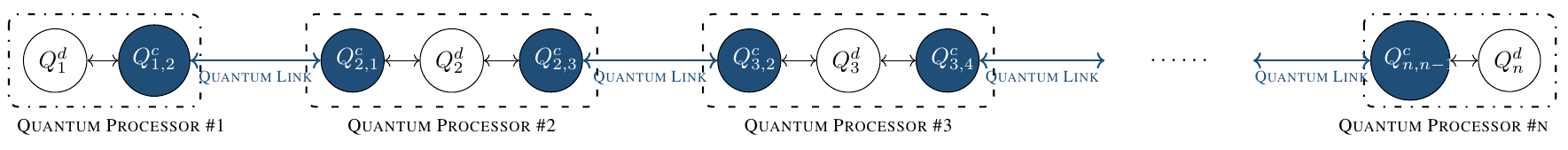}
	\end{adjustbox}
	\caption{Worst-case scenario in terms of overhead induced by the distributed computation: the quantum processors are interconnected through a one-dimensional nearest-neighbor topology, and only one data qubit is available at each quantum processor. Intra-processor coupling between communication qubits omitted for the sake of simplicity.}
	\label{Fig:07}
	\hrulefill
\end{figure*}

We consider the worst-case scenario shown in Figure~\ref{Fig:07}. More in detail, we assume that only one data qubit is available at each quantum processor\footnote{Clearly, the total number of data qubits within the distributed architecture must be greater than the number of logical qubits within the quantum circuit to be compiled.}. The rationale for this choice is as follows. Whenever multiple data qubits are available at a single quantum processor, a local \texttt{CNOT} can be executed between these data qubits without incurring in any overhead induced by the distributed computation. Conversely, with just one data qubit available at each processor, each and every \texttt{CNOT} within the quantum circuit must be mapped into a remote \texttt{CNOT}, and hence the overhead induced by the distributed computation is the highest possible.

Furthermore, we assume that the quantum processors are interconnected through a one-dimensional nearest-neighbor topology, as shown in Figure~\ref{Fig:07}. Again, the rationale for this choice is to consider the worst-case scenario in terms of overhead induced by the distributed computation. In fact, the considered topology is characterized by the lowest possible number of communications qubits -- i.e., $2 n - 2$ with $n$ denoting the number of quantum processors -- since the removal of any communication qubit would disconnect the network into two disjoint subsets of quantum processors. And the quantum processors are arranged in a line -- rather than in a star -- to maximize both the number of non-adjacent quantum processors and the maximum distance -- in terms of \textit{hops} -- between two non-adjacent quantum processors.

From the above, it becomes clear that the considered architecture represents the worst-case scenario in terms of overhead induced by the distributed computation. Hence, the \textit{actual}  overhead induced by any real-world architecture will be always upper-bounded by the communication overhead induced by the considered architecture.

Clearly, we need to choose a metric for measuring the overhead induced by the distributed computation. As discussed in Section~\ref{sec:2.2}, there exists a general consensus on circuit depth as a key performance metric of circuit compilation. Hence, in the following, we measure the overhead in terms of \textit{number of additional layers required to distribute the computation of a single layer in the original quantum circuit}. Furthermore, we also evaluate the overhead in terms of how many calls to the link entanglement generation process are required from the compiling algorithm.

\subsection{Basic Strategies for Distributing CNOTs}
\label{sec:4.2}

\begin{figure*}[ht!]
    \begin{minipage}[c]{.20\linewidth}
		\centering
        \begin{adjustbox}{width=\columnwidth}
            \includegraphics[]{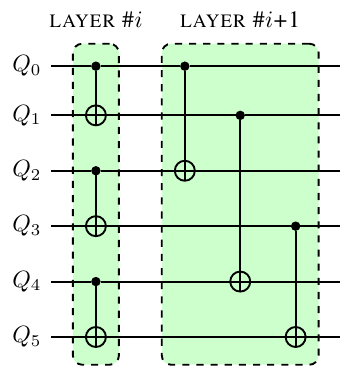}
        \end{adjustbox}
	    \subcaption{Original quantum circuit.}
        \label{Fig:08-a}
    \end{minipage}
    \hfil
    \begin{minipage}[c]{.68\linewidth}
    	\centering
    	\begin{adjustbox}{width=\columnwidth}
            \includegraphics[]{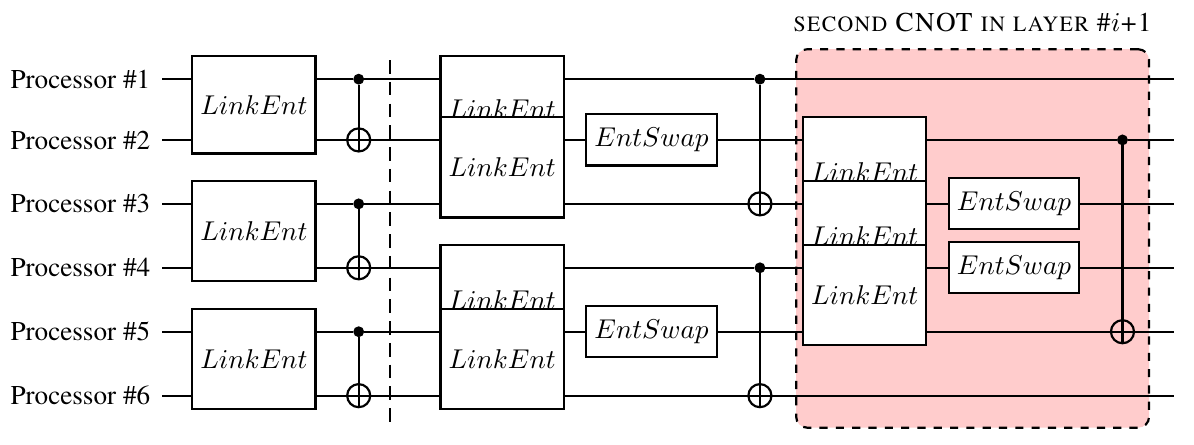}
        \end{adjustbox}
        \subcaption{Compiled quantum circuit.}
    	\label{Fig:08-b}
    \end{minipage}
    \caption{\textit{Entanglement swapping strategy}. Each remote \texttt{CNOT} in Figure~\ref{Fig:08-a} requires two preliminary tasks: i) \textit{Link Entanglement}, for distributing the entanglement between neighbor nodes, and, ii) the \textit{Entanglement Swapping}, for entangling the two remote processors involved within the \texttt{CNOT}. Clearly, the swapping task is omitted whenever the \texttt{CNOT} operates between data qubits stored at processors that are neighbor within the network topology, as for the $i$-th layer.}
    \label{Fig:08}
    \hrulefill
\end{figure*}

Let us consider a single layer of the original $n$-qubit quantum circuit. Clearly, the number of \texttt{CNOT}s in each layer is lower or equal to $\frac{n}{2}$, given that at most $\frac{n}{2}$ gates can be executed simultaneously (and thus belong to the same layer) by operating on different pairs of qubits.

As discussed in Section~\ref{sec:4.1}, we aim at considering the worst-case scenario in terms of overhead induced by the distributed computing architecture. Hence, each \texttt{CNOT} within the quantum circuit -- given that it operates on physical qubits stored at different processors as discussed in Section~\ref{sec:4.1} -- is a remote \texttt{CNOT}. As a consequence, the compiler must map at most $\frac{n}{2}$ remote \texttt{CNOT}s in each layer.

\subsubsection*{Entanglement Swapping Based Strategy}
The first strategy for implementing remote \texttt{CNOT}s is based on the \textit{entanglement swapping} technique discussed in Section~\ref{sec:3.1} and shown in Figure~\ref{Fig:06}.

Accordingly, each remote \texttt{CNOT} is implemented by firstly generating link entanglement \cite{Cal-17} among neighbor nodes. To this aim, different techniques for entanglement generation can be employed, depending on the particulars of the underlying qubit technology \cite{CacCalVan-20}. Nevertheless, link entanglements can be simultaneously generated, given that each processor is equipped with two communication qubits. Once generated, the entanglement is simultaneously\footnote{In general, the capability to generate (and to re-generate, once depleted) and distribute entangled Bell states  through different links in parallel depends on the quantum resources available, i.e., both the number of communication qubits at each processor and the inter-connection (shared bus vs. point-to-point) among the communication qubits. Differently, the possibility to simultaneously swap the entanglement at the intermediate nodes depends only on classical resources, i.e., the possibility to simultaneously transfer classical information.} swapped at intermediate nodes so that a Bell state is distributed between the two remote processors and, finally, the remote \texttt{CNOT} is obtained as shown in Figure~\ref{Fig:04-b}.

The \textit{entanglement swapping based strategy} is outlined in Figure~\ref{Fig:08-b} in terms of basic tasks. Within the figure, the particulars of each task are omitted for the sake of clarity. For instance, entanglement swapping -- although depicted as a single block -- is indeed obtained with a quantum circuit composed by three layers as shown in Figure~\ref{Fig:06-b}. Similarly, the link entanglement generation requires a quantum circuit with a depth equal or greater than two, depending on the particulars of the quantum technology underlying entanglement generation and distribution \cite{CacCalVan-20}.

Nevertheless, the figure\footnote{We note that -- for the sake of simplicity -- in Figure~\ref{Fig:08-b} we simply mapped the $j$-th logical qubit $Q_j$ of layer \#i in Figure~\ref{Fig:08-a} onto the $j+1$-th processor, ignoring so any optimization achievable with a proper mapping of the logical qubits of the quantum circuit onto the physical qubits of the quantum processor.} provides a clear intuition of both: i) the sequentiality constraints between the different tasks, and ii) the parallelism achievable within each task. Specifically, whenever the \texttt{CNOT}s \textit{overlaps}\footnote{The term ``\textit{overlap}'' indicates the case when the execution of the considered CNOTs involves overlapping sets of intermediate processors as a consequence of the constraint we imposed on the network topology of having $2n-2$ communication qubits. With reference to the example in Figure~\ref{Fig:08-a}, the CNOT between $Q_0$ and $Q_2$ in the layer \#i+1 overlaps with the CNOT between $Q_1$ and $Q_4$, being the communication qubits at the processors \#1 and \#2 needed to both of them. Differently, the CNOT between $Q_3$ and $Q_5$ does not overlap with CNOT between $Q_0$ and $Q_2$ and, hence, they can be performed in parallel.} within the network topology (as for the \texttt{CNOT}s of the layer \#i+1 in Figure~\ref{Fig:08-a}), they must be executed sequentially. Differently, \texttt{CNOT}s that don't overlap (as for the \texttt{CNOT}s of the $i$-th layer in Figure~\ref{Fig:08-a}) can be executed simultaneously. 
Since we are interested in assessing the worst-case overhead induced by distributed computation, in the following we consider the worst-case scenario in which all the \texttt{CNOT}s of an arbitrary layer of the quantum circuit \textit{overlap} within the network topology. Hence, we have that the depth overhead of the \textit{entanglement swapping based strategy} does not exceed the following depth:
\begin{equation}
    \label{eq:4.1}
    \frac{n}{2} \, d_{es}
\end{equation}
where $n$ denotes the number of logical qubits within the quantum circuit and $d_{es}$ is a constant factor (independent from the characteristics of the original quantum circuit) given by:
\begin{equation}
    \label{eq:4.2}
    d_{es} = c_{le} + c_{bsm} + c_{cx}
\end{equation}
with $c_{le}$ and $c_{bsm}$ denoting the number of layers required to perform the link entanglement task and the entanglement swapping task, respectively, and $c_{cx}$ denoting the number of layers required to perform a remote \texttt{CNOT} once the Bell state has been distributed between two processors. The actual values of $c_{le}$, $c_{bsm}$ and $c_{cx}$ depend on the particulars of the underlying hardware technology.

From \eqref{eq:4.1}, we have that the actual depth of an arbitrary $d$-depth quantum circuit compiled with the \textit{entanglement swapping based strategy} will be always lower than $\frac{n}{2} \, d$, neglecting the constant $d_{es}$. Hence, the depth overhead grows linearly with the number of logical qubits of the quantum circuit to be compiled. And given that this result holds for the worst-case scenario (one-data-qubit processors arranged in a one-dimensional network topology), the actual depth overhead induced by any arbitrary distributed architecture will be always upper-bounded by \eqref{eq:4.1}.

We further note that classical information must be exchanged between the quantum processors. For instance, the entanglement swapping task requires the transmission of classical information (i.e., the measurement output) throughout the quantum network. Hence, in case of long-distance quantum processors, the actual execution time of the compiled quantum circuit may be affected by the latency induced by the classical communications.

Finally, due to the complex and stochastic nature of the physical mechanisms underlying quantum entanglement \cite{Cal-17}, several attempts can be required for establishing a link entanglement, and this may impact as well the execution time of the compiled quantum circuit. Indeed, we should consider link entanglement as the \textit{critical task} for distributed quantum computation, given that the remaining tasks require only local quantum operations and classical communications. From this perspective, the \textit{entanglement swapping based strategy} requires at most $\frac{n}{2}$ repetitions of the link entanglement task, regardless of the original quantum circuit and regardless of the characteristics of the network topology underlying the distributed computing architecture.

\begin{figure*}[ht!]
    \begin{minipage}[c]{.31\linewidth}
		\centering
        \begin{adjustbox}{width=\columnwidth}
            \includegraphics[]{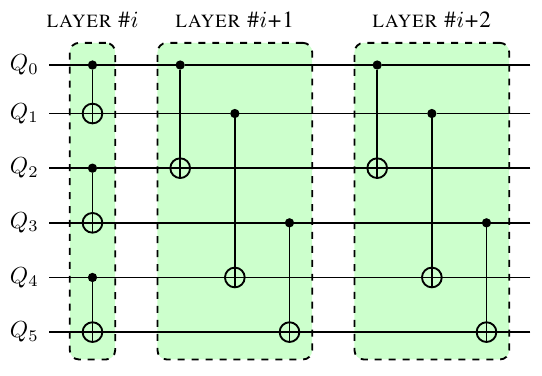}
        \end{adjustbox}
	    \subcaption{Original quantum circuit.}
        \label{Fig:09-a}
    \end{minipage}
    \hfil
    \begin{minipage}[c]{.67\linewidth}
    	\centering
    	\begin{adjustbox}{width=\columnwidth}
            \includegraphics[]{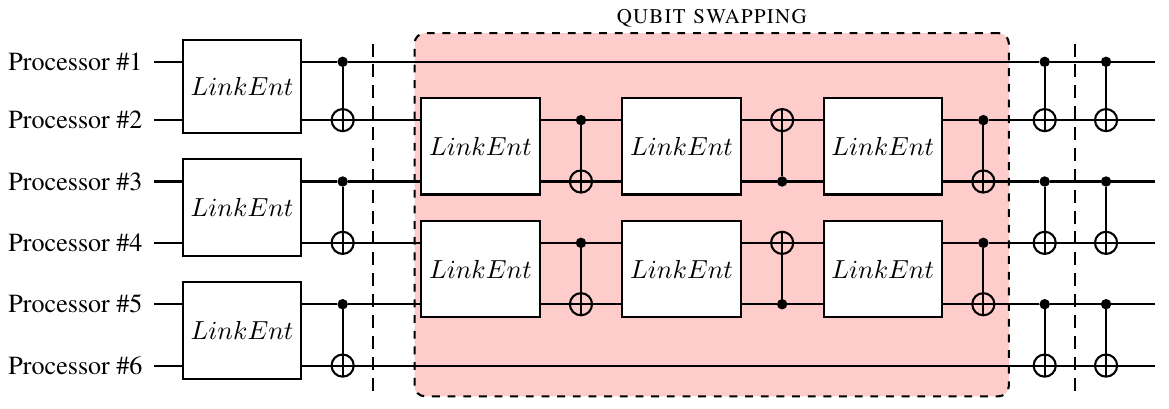}
        \end{adjustbox}
        \subcaption{Compiled quantum circuit.}
    	\label{Fig:09-b}
    \end{minipage}
    \caption{\textit{Data-qubit swapping based strategy}. Swapping data qubits between remote quantum processors can be advantageous whenever the original quantum circuit presents repetitions of the same \texttt{CNOT} interaction pattern, as for layers \#$i$+1 and \#$i$+2 in Figure~\ref{Fig:09-a}. Although not shown in the figure, the entanglement swapping tasks (as in Figure~\ref{Fig:08-b}) are needed whenever it is necessary to swap data qubits stored at processors that are not neighbor within the network topology.}
    \label{Fig:09}
    \hrulefill
\end{figure*}

\subsubsection*{Data-Qubit Swapping Based Strategy}

\begin{figure}[ht!]
    \begin{minipage}[c]{.33\linewidth}
		\centering
        \begin{adjustbox}{width=\columnwidth}
            \includegraphics[]{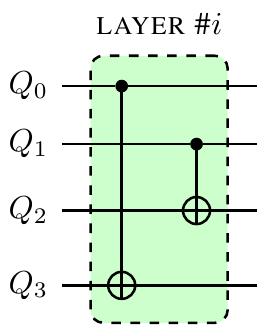}
        \end{adjustbox}
	    \subcaption{Original quantum circuit.}
        \label{Fig:10-a}
    \end{minipage}
    \hfil
    \begin{minipage}[c]{.32\linewidth}
    	\centering
    	\begin{adjustbox}{width=\columnwidth}
            \includegraphics[]{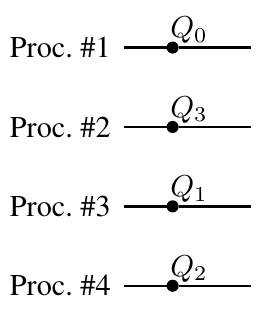}
        \end{adjustbox}
        \subcaption{Possible arrangement.}
    	\label{Fig:10-b}
    \end{minipage}
    \begin{minipage}[c]{.32\linewidth}
    	\centering
    	\begin{adjustbox}{width=\columnwidth}
            \includegraphics[]{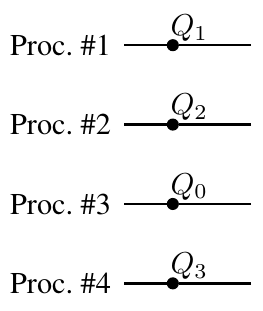}
        \end{adjustbox}
        \subcaption{Alternative equivalent arrangement.}
    	\label{Fig:10-c}
    \end{minipage}
    \caption{\textit{Data-qubit swapping based strategy}: equivalent mappings. Both Figure~\ref{Fig:10-b} and Figure~\ref{Fig:10-c} represent valid arrangements of the data qubits within the remote quantum processors so that each \texttt{CNOT} in Figure~\ref{Fig:10-a} operates on qubits stored at processors neighbor within the network topology.}
    \label{Fig:10}
    \hrulefill
\end{figure}

The \textit{entanglement swapping based strategy} takes full advantage of the augmented connectivity enabled by the communication qubits -- as discussed in Section~\ref{sec:3.1} -- to allow interactions between remote processors within each layer.

Nevertheless, whenever the original quantum circuit presents repetitions of the same \texttt{CNOT} interaction pattern between logical qubits -- as for layers \#$i$+1 and \#$i$+2 in Figure~\ref{Fig:09-a} -- a more elaborate strategy -- based on \textit{moving} the data qubits -- can provide better performance.

The strategy is shown in Figure~\ref{Fig:09}: the objective is \textit{to arrange} (i.e., to swap) the data qubits within the quantum processors so that eventually each \texttt{CNOT} of the original layer operates on qubits stored at neighbor processors within the network topology.

Intuitively, the strategy goal can be modeled as an array sorting problem. Indeed, similarly to classical sorting, the $n$ data qubits (representing the \textit{values} to be sorted) must be ordered within the network topology (representing an array with size equal or greater than $n$ ). However, differently from classical sorting where any couple of values can be swapped regardless from their position within the array, with the \textit{data-qubit swapping} the constraints arising from the underlying network topology must be carefully taken into account. To this aim, by taking advantage of the \textit{sorting network theory}, it is easy to model the network topology constraints through the notion of \textit{insertion network} (or, equivalently, \textit{bubble network}). As a consequence, the overall depth of the equivalent quantum circuit grows with the number $n$ of logical qubits as \cite{Knu-98}:
\begin{equation}
    \label{eq:4.3}
    2 n - 3
\end{equation}
instead of a logarithmic $\log{n}$ depth factor as for classical sorting.

\begin{algorithm}[t]
    \caption{Data-Qubit Swapping
        \newline
        \footnotesize
        \textbf{Input}: $n$-qubit circuit layer $L$ with mod(n,4) = 0 and $\frac{n}{2} \texttt{CNOT}s$
        \newline
        \textbf{Output}: layer $L$ with each \texttt{CNOT} operating on neighbor qubits
    }
    \label{Alg:01}
    \begin{algorithmic}[1]
        \Function{Sort}{$L$}
            \If{$\exists \, \texttt{CNOT}(q_i,q_j) \text{ with } \; i,j \leq \frac{n}{2}$} \label{Alg:01-1}
                \State \Comment $\exists \, \texttt{CNOT}(q_k,q_l) \text{ with } k,l > \frac{n}{2}$ 
                \State \Call{Swap}{$q_{i+1},q_j$} \label{alg:1.1.a}
                \State \Call{Swap}{$q_{k+1},q_l$}
                \State $L = L \setminus \{q_i,q_{i+1},q_k,q_{k+1}\}$
            \Else
                \State \Comment $\exists \, \texttt{CNOT}(q_{\frac{n}{2}},q_l) \text{ with } l > \frac{n}{2}$ 
                \State \Comment and $\exists \, \texttt{CNOT}(q_i,q_{l-1}) \text{ with } i < n/2$
                \State \Call{Swap}{$q_{\frac{n}{2}},q_{l-1}$}
                \State \Call{Swap}{$q_i,q_{\frac{n}{2}-1}$}  
                \State $L = L \setminus \{q_{\frac{n}{2}-1},q_{\frac{n}{2}},q_{l-1},q_l\}$
            \EndIf
            \If{$L \neq \emptyset$} 
                \State \Call{Sort}{$L$}
            \EndIf
        \EndFunction
    \end{algorithmic}
\end{algorithm}

Nevertheless, sorting networks -- and in general classical sorting -- are based on the assumption that there exists a total (monotonic) order over the array elements. Hence, there exists a unique solution to the sorting problem. Conversely, the  \textit{data-qubit swapping based strategy} admits several equivalent solutions for the arranging problem, as exemplified in Figure~\ref{Fig:10}. We now formalize these considerations with the following theorem.

\begin{theo}
    \label{Theo:01}
    Let us consider the $i$-th layer of an arbitrary $n$-qubit quantum circuit. The depth of the corresponding compiled quantum circuit, obtained through the \textit{data-qubit swapping based strategy}, does not exceed the following depth:
    \begin{equation}
        \label{eq:4.4}
        \frac{n}{4} \, d_{qs} + d'_{qs}
    \end{equation}
    where $d_{qs}$ and $d'_{qs}$ are constant factors (independent from the characteristics of the original quantum circuit) given by:
    \begin{align}
        \label{eq:4.5}
        d_{qs} = &3 \left( c_{le} + c_{bsm} + c_{cx} \right)\\
        \label{eq:4.6}
        d'_{qs} &= c_{le} + c_{cx}
    \end{align}
    with $c_{le}$ and $c_{bsm}$ denoting the number of layers required to perform the link entanglement task and the entanglement swapping task, respectively, and $c_{cx}$ denoting the number of layers required to perform the remote \texttt{CNOT}s once the Bell state has been distributed between two processors.
    \begin{proof}
    The proof easily follows by recognizing that: i) the function $\Call{SORT}{\cdot}$, defined in Algorithm~\ref{Alg:01}, is called at most $\frac{n}{4}$ times, and ii) after these calls to $\Call{SORT}{\cdot}$, all the \texttt{CNOT}s, by acting on qubits stored at neighbor processors, can be executed at once through link entanglement followed by local operations, as shown in Figure~\ref{Fig:04-b}.\\ 
    More specifically, in each call we have two disjoint cases (line 2).\\
    In the former case (lines 3-6), there exists a \texttt{CNOT} acting within the first \textit{half portion} -- i.e., the first $\frac{n}{2}$ logical qubits -- of the original layer. Since we are considering the worst-case -- namely, a layer with $\frac{n}{2}$ \texttt{CNOT}s -- then we have that there exists at least one \texttt{CNOT} acting on the last \textit{half portion} -- i.e., the last $\frac{n}{2}$ logical qubits -- of the original layer. Hence, the two \texttt{CNOT}s do not overlap and two simultaneous \texttt{SWAP} operations can be executed -- one in each \textit{half portion} of the original quantum circuit -- as shown here:
    \begin{equation}
        \begin{tikzpicture}
            \matrix  (m) [matrix of math nodes, nodes={inner sep=1pt}, column sep=0.2ex, row sep=4ex] {
                q_1 & \ldots & q_i & q_{i+1} & \ldots & q_j & \ldots & q_{\frac{n}{2}} & \ldots & q_k & q_{k+1} & \ldots & q_l & \ldots q_n \\
                q_1 & \ldots & q_i & q_j & \ldots & q_{i+1} & \ldots & q_{\frac{n}{2}} & \ldots & q_k & q_l & \ldots & q_{k+1} & \ldots q_n \\                
            };
            \draw[->]
                (m-1-4) edge (m-2-6)
                (m-1-6) edge (m-2-4)
                (m-1-11) edge (m-2-13)
                (m-1-13) edge (m-2-11);
        \end{tikzpicture} \nonumber
    \end{equation}
    so that, within the compiled circuit -- the two \texttt{CNOT}s act on qubits stored at neighbor processors. Within the previous diagram, as well as in Algorithm~\ref{Alg:01}, we omitted some minor particulars for the sake of simplicity. For instance, we implicitly assumed that $i$ (and $k$) is odd -- i.e., $mod(i,2) = 1$ -- so that $q_j$ must be swapped with $q_{i+1}$. Clearly whether $i$ should be even, $q_j$ must be swapped with $q_{i-1}$.\\
    In the latter case (lines 8-12), each and every \texttt{CNOT} acts on two logical qubits belonging to both the \textit{half portions} of the original layer. Let us consider, with no lack of generality, the \texttt{CNOT} acting on the $\frac{n}{2}$-th logical qubit (i.e., the last qubit of the first \textit{half portion}) and let us denote as $q_l$ the second qubit on which such a \texttt{CNOT} operates. Since we are considering a layer with $\frac{n}{2}$ \texttt{CNOT}s, then we have that there exists a \texttt{CNOT} acting on the $l-1$-th qubit. By denoting as $q_i$ the second qubit on which such a \texttt{CNOT} operates, it follows $i < \frac{n}{2}$. Although the two \texttt{CNOT}s overlap, by properly selecting two \texttt{SWAP} operations as shown here:
        \begin{equation}
        \begin{tikzpicture}
            \matrix  (m) [matrix of math nodes, nodes={inner sep=1pt}, column sep=1ex, row sep=4ex] {
                q_1 & \ldots & q_i & \ldots & q_{\frac{n}{2}-1} & q_{\frac{n}{2}} & \ldots & q_{l-1} & q_{l} & \ldots q_n \\
                q_1 & \ldots & q_{\frac{n}{2}-1} & \ldots & q_i & q_{l-1} & \ldots & q_{\frac{n}{2}} & q_{l} & \ldots q_n \\
            };
            \draw[->]
                (m-1-6) edge (m-2-8)
                (m-1-8) edge (m-2-6)
                (m-1-3) edge (m-2-5)
                (m-1-5) edge (m-2-3);
        \end{tikzpicture} \nonumber
    \end{equation}
    we have that the \texttt{SWAP}s can be simultaneously executed so that, within the compiled circuit -- the two \texttt{CNOT}s act on qubits stored at neighbor processors. Regardless of which case holds, each call to the $\Call{SORT}{\cdot}$ function compiles two \texttt{CNOT}s. By recalling that at most $\frac{n}{2}$ \texttt{CNOT}s are present in a layer, the thesis follows.
    \end{proof}
\end{theo}

From \eqref{eq:4.4}, we have that the actual depth of an arbitrary $d$-depth quantum circuit compiled with the \textit{data-qubit swapping based strategy} will be always lower than $\frac{n}{4} \, d$, neglecting the constant factors. Hence, the depth overhead is asymptotically lower than the overhead induced by the \textit{entanglement swapping based strategy}. However, an explicit comparison between the two strategies depends on the particulars of the underlying qubit technology through the exact expressions of $d_{es}$ and $d_{qs}$. Furthermore, it depends also on the repetitions of the same \texttt{CNOT} interaction patterns within the original quantum circuit.

As regards to the number of repetitions of the link entanglement task, in general it depends on the characteristics of the underlying qubit technology as discussed in Section~\ref{sec:4.3}. With reference to the IBM quantum processors, where a \texttt{SWAP} operation is obtained through a sequence of three \texttt{CNOT}s, from \eqref{eq:4.4}-\eqref{eq:4.6} we have that the \textit{data-qubit swapping based strategy} requires at most $2 n$ repetitions of the link entanglement task, regardless of the original quantum circuit and regardless of the characteristics of the network topology underlying the distributed computing architecture.

\subsection{Discussion}
\label{sec:4.3}

As already mentioned above, the performance of the two strategies firmly depends on the particulars of the underlying hardware technology through the parameters $d_{es}$, $d_{qs}$ and $d'_{qs}$ given in equations \eqref{eq:4.2}, \eqref{eq:4.5} and \eqref{eq:4.6}.

To better clarify this point, let us consider $d_{qs}$, which inherently denotes the cost for a \texttt{SWAP} operation. Having assumed through the manuscript the \texttt{CNOT} being the fundamental multi-qubit gate, a single remote \texttt{SWAP} operation can be obtained through three remote \texttt{CNOTs} as in Figure~\ref{Fig:03-b}. And this is the rationale for the constant factor equal to $3$ in \eqref{eq:4.5}, which accounts for the cost of three remote \texttt{CNOTs}.

Clearly, by changing the assumptions on the underlying hardware technology, the expression of $d_{qs}$ changes as well. For instance, photonic technology can provide the \texttt{SWAP} gate as the native operation \cite{OnoOkaTan-17}, and in such a case $d_{qs}$ is equal to $1$. Nevertheless the main result -- i.e., equation \eqref{eq:4.4} -- continues to hold. Furthermore, whenever the \texttt{SWAP} gate is the native operation, a single \texttt{CNOT} can be obtained through two consecutive \texttt{SWAP}s interleaved by single-qubit operations \cite{SchSie-03}. Hence, the expression of $d_{es}$ must change accordingly but the main result -- i.e., equation \eqref{eq:4.2} -- continues to hold as well.

Indeed it is worthwhile to note that, despite the differences between the performance of the two strategies, there exists a one-to-one mapping between the strategies. Specifically, there exists an admissible transformation allowing to map the compiled circuit obtained with a strategy into the compiled circuit obtained with the other strategy. And the corresponding computational task exhibits a polynomial-time complexity\footnote{Given that both the strategies exhibit a polynomial-time computational complexity as proved in Section~\ref{sec:5.1}.} for every original circuit.

\section{Performance Analysis}
\label{sec:5}

Here we perform a performance analysis for the compiler design conducted in Section~\ref{sec:4}.

More in detail, in Section~\ref{sec:5.1} we illustrate the algorithmic implementation of the compiler, proving so its attractive feature -- \textit{a polynomial time complexity that grows polynomially with the number of logical qubits and linearly with the depth of the quantum circuit to be compiled} -- from a computational perspective. Then, in Section~\ref{sec:5.2} we validate the theoretical upper bounds on the number of layers that result from compiling a layer of remote CNOTs, derived in Section~\ref{sec:4.2}, against an extensive set of medium-size quantum circuits of practical interest. Finally, with Sections~\ref{sec:5.3} and \ref{sec:5.4} we conclude the performance analysis through an \textit{unfair} -- as clearly shown with Figure~\ref{Fig:hypercube} -- comparison with the state-of-the art for two different network topologies.

\begin{figure*}[ht!]
     \centering
     \includegraphics[width=15cm]{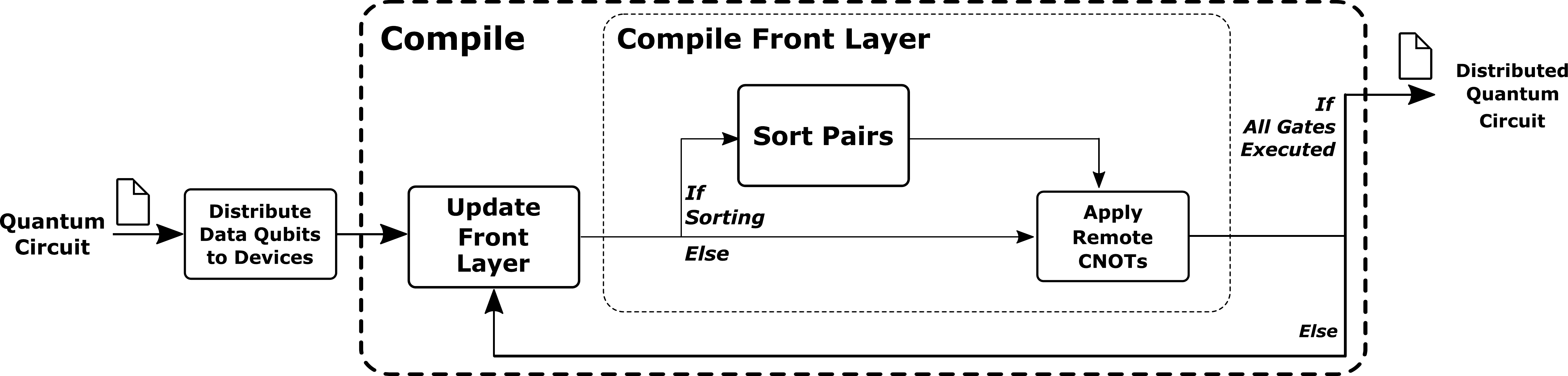}
     \caption{Workflow of the proposed compiler.}
     \label{Fig:compiler_workflow}
\end{figure*}

\subsection{Compiler Implementation}
\label{sec:5.1}

We implemented the strategies discussed in Section~\ref{sec:4.2} in Python, using Qiskit \cite{Qiskit} as the development framework. Given a quantum circuit described in the QASM format, the compiler proceeds to instantiate a distributed architecture that mimics the one described in Section~\ref{sec:4.1}. To model the worst-case scenario depicted in Figure~\ref{Fig:07}, each QPU has one data qubit and two communication qubits. Moreover, each QPU has two neighbor QPUs, with the exception of the outer QPUs that have one neighbor QPU only. Each qubit of the circuit is assigned to the data qubit of one QPU.

\begin{algorithm}[H]
    \caption{\textsc{CompileFrontLayer}\newline
    \footnotesize
    \textbf{Input}: the front layer $\mathcal{F}$ to be compiled, the executed gates $\mathcal{E}$ until now\newline
    \textbf{Output}: the executed gates $\mathcal{E}$ updated with the compiled front layer $\mathcal{F}$}
    \label{alg:compile_front}
    \begin{algorithmic}
    \footnotesize
	\If{\textit{data-qubit swapping based strategy}}
        \State prepare $interactions$ vector
        \State $swaps \gets \textsc{SortPairs(interactions)}$
        \ForAll{$swap \in swaps$} \Comment{Perform remote SWAPs}
            \If{two EPR pairs between QPUs}
                \State \Call{Teleport}{$swap.q1, swap.q2$}
            \Else
                \State \Call{RemoteCnot}{$swap.q1, swap.q2$}
                \State \Call{RemoteCnot}{$swap.q2, swap.q1$}
                \State \Call{RemoteCnot}{$swap.q1, swap.q2$}
            \EndIf
        \EndFor
    \EndIf
    \ForAll{$gate \in \mathcal{F}$}
        \State \Call{RemoteCnot}{$gate.control, gate.target$}
    \EndFor
    \State \textbf{return} $\mathcal{E}$
	
	\Statex
	\Function {RemoteCnot}{$control, target$}
	\footnotesize
	    \If{$control$ and $target$ are not on neighboring QPUs}
	        \State $qpu_0 \gets$ QPUs of $control$
            \State $qpu_n \gets$ QPUs of $target$
            \For{$i \in \{0,..,n-2\}$}
            \State add \textit{link entanglement}
            \State \hspace{\algorithmicindent}between $qpu_i$ and $qpu_{i+1}$ to $\mathcal{E}$
            \EndFor
            \State add \textit{entanglement swap} between $qpu_0$ and $qpu_n$ to $\mathcal{E}$
        \EndIf
        \State add \textit{link entanglement} between $qpu_0$ and $qpu_n$ to $\mathcal{E}$
        \State add a remote \textsc{CNOT} between $control$ and $target$ to $\mathcal{E}$
	\EndFunction
	\Statex
	\Function {Teleport}{$q1, q2$}
	\footnotesize
	    \If{$q1$ and $q2$ are not on neighboring QPUs}
	        \State $qpu_0 \gets$ QPUs of $q1$
            \State $qpu_n \gets$ QPUs of $q2$
            \For{$i \in \{0,..,n-2\}$}
                \State add \textit{link entanglement}
                \State \hspace{\algorithmicindent}between $qpu_i$ and $qpu_{i+1}$ to $\mathcal{E}$
                \State \hspace{\algorithmicindent}using two available EPR pairs between QPUs
            \EndFor
            \State $e_{01}, e_{02} \gets $ the two communication qubits at $qpu_0$
            \State $e_{n1}, e_{n2} \gets $ the two communication qubits at $qpu_n$
            \State add \textit{entanglement swap} between $e_{01}$ at $qpu_0$ and
            \State \hspace{\algorithmicindent}$e_{n1}$ at $qpu_n$ to $\mathcal{E}$
            \State add \textit{entanglement swap} between $e_{02}$ at $qpu_0$ and
            \State \hspace{\algorithmicindent}$e_{n2}$ at $qpu_n$ to $\mathcal{E}$
            \State add \textit{quantum teleportation} between $q1$ and $e_{n1}$ to $\mathcal{E}$
            \State add \textit{quantum teleportation} between $q2$ and $e_{02}$ to $\mathcal{E}$
            \State add local \textit{swap} between $q1$ and $e_{02}$
            \State add local \textit{swap} between $q2$ and $e_{n1}$
	    \EndIf
	\EndFunction
    \end{algorithmic}
\end{algorithm}

The compilation process is summarized and illustrated in Figure~\ref{Fig:compiler_workflow}. 
Reading the circuit from left to right, the \textit{front layer}, \textit{i.e.}, a layer comprising only \texttt{CNOT} gates that can be executed in parallel, is updated. To this end, one-qubit gates are immediately mapped to the compiled circuit, while CNOT gates are added to the front layer. This is done until all logical qubits are interested by a CNOT or there are no more CNOT gates that can be executed in parallel in the current front layer. Then, the front layer is compiled, rendering all currently involved CNOT gates executable. This process is repeated until no more front layers can be computed, meaning that all the circuit gates have been mapped to the distributed architecture.

\begin{algorithm}[h!]
	\caption{
		\textsc{SortPairs}\newline
		\footnotesize
		\textbf{Input}: a vector \textit{v} representing CNOT interactions between qubits pairs, ex. [1 2 2 3 1 3]\newline
		\textbf{Output}: the $swaps$ to perform}
	\label{alg:sort}
	\begin{algorithmic}
		\footnotesize
		\State \Comment{\textbf{\textit{f.e.o.}} stands for 'first element of'}
		\If{length($v$) $\bmod 4 \neq 0$} \Comment{\textit{length(v)} must be multiple of 4}
			\State add $\{-1, -1\}$ to $v$ \Comment{Add a dummy pair at the end}
		\EndIf
		\State $swaps \gets \emptyset$ \Comment{List of SWAPs to perform}
		\State $n \gets$ length($v$)
		\State $mask1 \gets \{0, ..., \frac{n}{2}-1\}$
		\State $mask2 \gets \{\frac{n}{2}, ..., n-1\}$
		\State $cycle \gets 0$
		\While{$cycle < n/4$}
			\State $w \gets$ indexes of unique values in $v$
			\State $index\_last \gets \{0, ..., |v[mask1]|\} \setminus w$
			\If{$index\_last = \emptyset$}
				\State swap $v[mask1[end]]$ and
				\State \hspace{\algorithmicindent}\textbf{\textit{f.e.o.}} s.t. $v[mask2] \neq v[mask1[end]]$
				\State update $swaps$
			\EndIf
			\State $v, mask1, swaps \gets$ \Call{swap}{$v$, $mask1$, $swaps$}
			\State $v, mask2, swaps \gets$ \Call{swap}{$v$, $mask2$, $swaps$}
		\EndWhile
		\State \textbf{return} $swaps$
		
		\Statex
		\Function {swap}{$v,mask, swaps$}
		\footnotesize
			\State $w \gets$ indexes of unique values in $v$
			\State $index\_last \gets$ \textbf{\textit{f.e.o.}} $\{0, ..., |v[mask]|\} \setminus w$
			\State $index\_first \gets$ \textbf{\textit{f.e.o.}} $v[mask]$ s.t. = $v[mask[index\_last]]$
			\If{$index\_last$ is odd}
			\State $index\_swap \gets index\_last - 1$
			\Else
			\State $index\_swap \gets index\_last + 1$
			\EndIf
			\State swap $v[mask[index\_swap]]$ and $v[mask[index\_first]]$
			\State update $swaps$
			\State remove $mask[index\_swap]$ and $mask[index\_last]$ from $mask$
			\State \textbf{return} $v, mask, swaps$
		\EndFunction
	\end{algorithmic}
\end{algorithm}

Front layer compilation is based on Algorithm~\ref{alg:compile_front}, where one can choose between the two strategies discussed in Section~\ref{sec:4.2}. When the \textit{entanglement swapping based strategy} is adopted, remote \texttt{CNOT} gates preceded by entanglement swapping are applied whenever the involved QPUs are not neighbors. Note that to perform entanglement swapping, as well as remote \texttt{CNOT}s, we need to generate link entanglement between all involved QPUs.

\begin{figure*}[h!]
\begin{minipage}[c]{0.17\linewidth}
	\centering
	\begin{adjustbox}{width=1\linewidth}
        \includegraphics[]{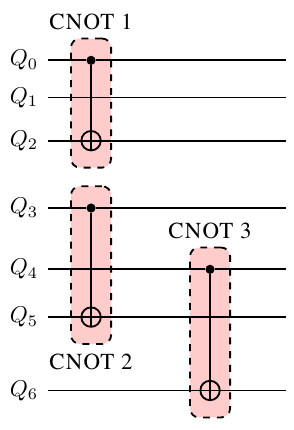}
	\end{adjustbox}
	\subcaption{A layer with three parallel CNOTs.}
	\label{Fig:sort_layer}
\end{minipage}
\begin{minipage}[c]{0.32\linewidth}
	\centering
	\begin{adjustbox}{width=1\linewidth}
        \includegraphics[]{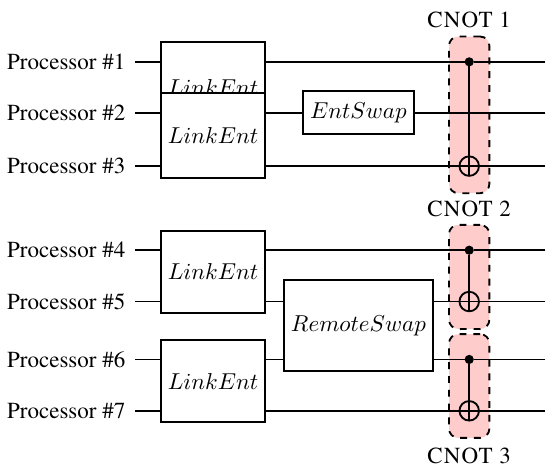}
	\end{adjustbox}
	\subcaption{The layer distributed with the Sort strategy.}
	\label{Fig:sort_layer_swap}
\end{minipage}
\begin{minipage}[c]{0.5\linewidth}
	\centering
	\begin{adjustbox}{width=1\linewidth}
        \includegraphics[]{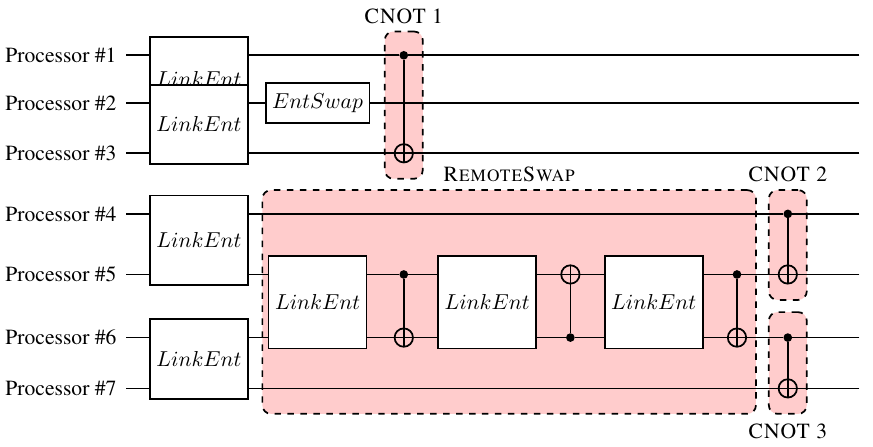}
	\end{adjustbox}
	\subcaption{The distributed layer after decomposing the Remote SWAP.}
	\label{Fig:sort_layer_decomp}
\end{minipage}
\caption{Compilation of a layer with three parallel CNOTs using the sorting strategy.}
\label{Fig:sort_explosion}
\hrulefill
\end{figure*}

Regarding the more advanced \textit{data-qubit swapping based strategy}, the list representing the interaction between qubits is prepared and then Algorithm~\ref{alg:sort} is used to compute the data swap operations needed to reorder the qubits. Referring to Figure~\ref{Fig:sort_layer}, the list representing the interaction between qubits before swapping would be $[1 1 2 3 2 3]$, while the sorted list after swapping would be $[1 1 2 2 3 3]$, meaning that no overlapping \texttt{CNOT}s are left in the layer.

The \textit{data-qubit swapping} routine operates on lists with a number of elements -- i.e., a number of logical qubits -- that is a multiple of 4. For this reason, a couple of dummy values, set to $-1$, may have to be added to the end of the list whenever the number of \texttt{CNOT}s is odd. Given that the algorithm searches for swaps form left to right, the dummy couple at the end will be left untouched. After creating \textit{masks} to keep track of already swapped qubits, the algorithm finds all necessary swaps in exactly $\frac{n}{4}$ steps, where $n$ is the number of elements in the list, i.e., the number of logical qubits interested by CNOTs.

Taking into account the topology of the worst-case scenario shown in Figure~\ref{Fig:07}, to perform a swap at least three remote CNOTs are needed.
After all the data-qubit swaps have been applied, the necessary remote CNOTs have to be placed. Ideally, this last step would involve only remote CNOTs between neighbor QPUs, but when not all qubits of the circuit are involved in the front layer, such as $Q_1$ in Figure~\ref{Fig:sort_layer}, it may be still necessary to perform some entanglement swapping operations, as shown in Figure~\ref{Fig:sort_layer_swap}. This is because, when sorting pairs, our algorithm does not take into account QPUs that are not involved in the current front layer. Consequently, the implementation of the \textit{data-qubit swapping based strategy} is actually an hybrid between the two strategies described in Sections~\ref{sec:4.2}, avoiding data-qubit swapping if not necessary and resorting instead to entanglement swapping.

\begin{figure*}[h!]
	\centering
	\begin{adjustbox}{width=1\linewidth}
        \includegraphics[]{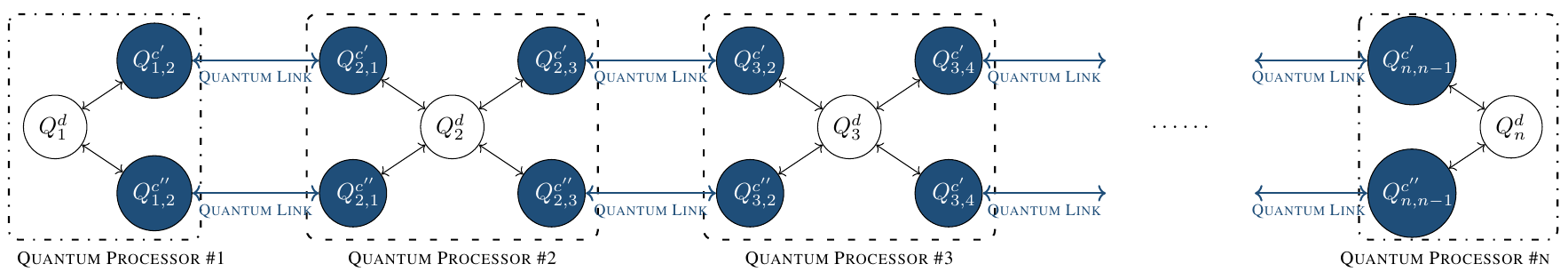}
	\end{adjustbox}
	\caption{Improving over the worst-case scenario topology. Only one data qubit is available at each quantum processor, but neighboring quantum processors are connected with two quantum links, realized by 2 EPR pairs.}
	\label{Fig:12}
	\hrulefill
\end{figure*}

As shown in Figure~\ref{Fig:sort_explosion}, starting from the layer in Figure~\ref{Fig:sort_layer}, following the \textit{data-qubit swapping based strategy}, in Figure~\ref{Fig:sort_layer_swap} we apply a remote SWAP between qubit 3 and qubit 4, and then we can execute all remote CNOTs in parallel.

Moving from the topology illustrated in Figure~\ref{Fig:07} to the one in Figure~\ref{Fig:12}, we can devise a different strategy to execute swaps by exploiting the augmented connectivity, described in Algorithm~\ref{alg:compile_front}.
Specifically, to perform a SWAP between QPU $Q_x$ and $Q_y$, our compiler uses one communication qubit at each QPU as a buffer memory and exploits quantum teleportation to move the state of a data qubit from $Q_x$ to $Q_y$ and vice versa, in parallel. After the two parallel teleportations, we can execute two parallel local SWAPs between communication qubits and data qubits at each QPU to effectively achieve data qubit swapping between $Q_x$ and $Q_y$. This is clearly beneficial as it only requires one layer of link entanglement generation between the interested QPUs, unlike the previous scenario where we needed three layer of link entanglement generation to perform three remote CNOTs.

The difference between \textit{data-qubit swapping} and \textit{entanglement swapping} lies in the fact that, if the subsequent front layers are similar (in terms of \texttt{CNOT} interaction pattern) to the one just compiled, not much data swapping and very little entanglement swapping (given that the front layers involve most of the qubits) will be necessary to compile those layers.

Regarding the computational complexity, Alg.~\ref{alg:compile} reads the circuit from left to right while updating the front layer, so its computational complexity is $O(d)$, where $d$ is the depth of the circuit, i.e., the number of layers. Alg.~\ref{alg:compile_front} loops through all necessary swaps found by Alg.~\ref{alg:sort}, and applies remote \texttt{CNOT}s. As we must take into account for possible \textit{entanglement swapping} operations between QPUs, applying a remote \texttt{CNOT} has a computational complexity of $O(n)$, with $n$ being the number of QPUs. Given that the we need at most $n/4$ swaps, the computational complexity of Alg.~\ref{alg:compile_front} turns out to be $O(n^2)$. The overall computational complexity of distributing a circuit is therefore $O(dn^2)$.

\begin{table*}[th!]
    \hrulefill
    \centering
    \resizebox{\textwidth}{!}{
    {
    \begin{tabular}{ | l | l | l | l | l | l | l | l | l |}
    \hline
    \multirow{ 2}{*}{circuit name} & \multirow{ 2}{*}{\# qubits $n$} & \multirow{ 2}{*}{\# CNOT layers} & \multicolumn{3}{c|}{Entanglement Swap} & \multicolumn{3}{c|}{Data-Qubit Swap}\\
    \cline{4-9}
    &  &  & \# CNOT layers $\times \frac{n}{2}d_{es}$  & compiled \# layers & compiled/theoretical & \# CNOT layers $\times (\frac{n}{4}d_{qs} + d'_{qs})$ & compiled \# layers & compiled/theoretical\\ \hline
	4gt12-v1\_89 & 16 & 88 & 2112 & 212 & 0.10 & 3344 & 233 & 0.07 \\ \hline
	4gt4-v0\_73 & 16 & 160 & 3840 & 372 & 0.10 & 6080 & 399 & 0.07 \\ \hline
	4mod7-v1\_96 & 16 & 65 & 1560 & 151 & 0.10 & 2470 & 155 & 0.06 \\ \hline
	9symml\_195 & 16 & 12849 & 308376 & 34721 & 0.11 & 488262 & 32809 & 0.07 \\ \hline
	adder & 10 & 55 & 825 & 144 & 0.17 & 1100 & 187 & 0.17 \\ \hline
	alu-v2\_31 & 16 & 172 & 4128 & 459 & 0.11 & 6536 & 436 & 0.07 \\ \hline
	ghz\_20 & 20 & 19 & 570 & 56 & 0.10 & 893 & 56 & 0.06 \\ \hline
	ghz\_4 & 4 & 3 & 18 & 8 & 0.44 & 33 & 8 & 0.24 \\ \hline
	H2\_RYRZ & 4 & 25 & 150 & 66 & 0.44 & 275 & 69 & 0.25 \\ \hline
	H2\_UCCSD & 4 & 52 & 312 & 72 & 0.23 & 572 & 72 & 0.13 \\ \hline
	H2O\_RYRZ & 14 & 125 & 2625 & 971 & 0.37 & 3625 & 2040 & 0.56 \\ \hline
	H2O\_UCCSD & 14 & 12937 & 271677 & 16017 & 0.06 & 375173 & 16017 & 0.04 \\ \hline
	ising\_model\_16 & 16 & 20 & 480 & 31 & 0.06 & 760 & 31 & 0.04 \\ \hline
	life\_238 & 16 & 8356 & 200544 & 22391 & 0.11 & 317528 & 21073 & 0.07\\ \hline
	LiH\_RYRZ & 12 & 105 & 1890 & 711 & 0.38 & 3045 & 1547 & 0.51 \\ \hline
	LiH\_UCCSD & 12 & 7264 & 130752 & 9216 & 0.07 & 210656 & 9216 & 0.04 \\ \hline
	one-two-three-v2\_100 & 16 & 29 & 696 & 76 & 0.11 & 1102 & 69 & 0.06 \\ \hline
	Random\_20q\_RYRZ & 20 & 185 & 5550 & 1991 & 0.36 & 8695 & 4113 & 0.47 \\ \hline
	Random\_20q\_UCCSD & 20 & 110497 & 3314910 & 130197 & 0.04 & 5193359 & 130197 & 0.03 \\ \hline
	random1\_n5\_d5 & 5 & 15 & 90 & 69 & 0.77 & 165 & 53 & 0.32\\ \hline
	random2\_n16\_d16 & 16 & 48 & 1152 & 666 & 0.58 & 1824 & 588 & 0.32 \\ \hline
	rd53\_138 & 16 & 42 & 1008 & 116 & 0.12 & 1596 & 100 & 0.06 \\ \hline
	root\_255 & 16 & 5965 & 143160 & 16699 & 0.12 & 226670 & 15973 & 0.07 \\ \hline
	sqn\_258 & 16 & 3719 & 89256 & 9713 & 0.11 & 141322 & 9210 & 0.07 \\ \hline
	sym9\_146 & 16 & 91 & 2184 & 277 & 0.13 & 3458 & 254 & 0.07 \\ \hline
    \end{tabular}}}
    \caption{Validation of the theoretical upper bounds derived in Section~\ref{sec:4.2} against a heterogeneous set of quantum circuits. Each circuit is characterized by a number of qubits $n$ and a number of CNOT layers, i.e., layers that comprise only CNOT gates. For each compiling strategy, there is a theoretical upper bound on the number of layers that are necessary to realize the remote CNOTs and an actual number of layers resulting from the compilation process. The ratio between the latter and the former is also reported.}
    \label{tab:overhead}
\end{table*}

\subsection{Compiling Overhead Validation}
\label{sec:5.2}

We validate the theoretical upper bounds (derived in Section~\ref{sec:4.2}) on the number of layers that result from compiling a layer of remote CNOTs, considering an extensive set of medium-size quantum circuits (the largest ones requiring 16 qubits, with the exception of a GHZ and two random circuit with 20 qubits). Specifically, we consider quantum circuits that are publicly available and widely adopted for testing quantum compilers~\cite{ZulPalWil-19,Li2019}\footnote{\url{https://github.com/deeptechlabs/quantum_compiler_optim/tree/master/examples}}, plus a few quantum chemistry circuits for the implementation of the unitary quantum Coupled-Cluster~\cite{Peruzzo2014,Barkoutsos2018} (qUCC) and the RYRZ heuristic~\cite{Kandala2017} wavefunction Ans\"atze.

\begin{figure*}[ht!]
    \centering
    \subfloat[Linear Nearest Neighbor topology.]{\label{Fig:hypercube_lnn}
    \fbox{
    \includegraphics[width=5cm]{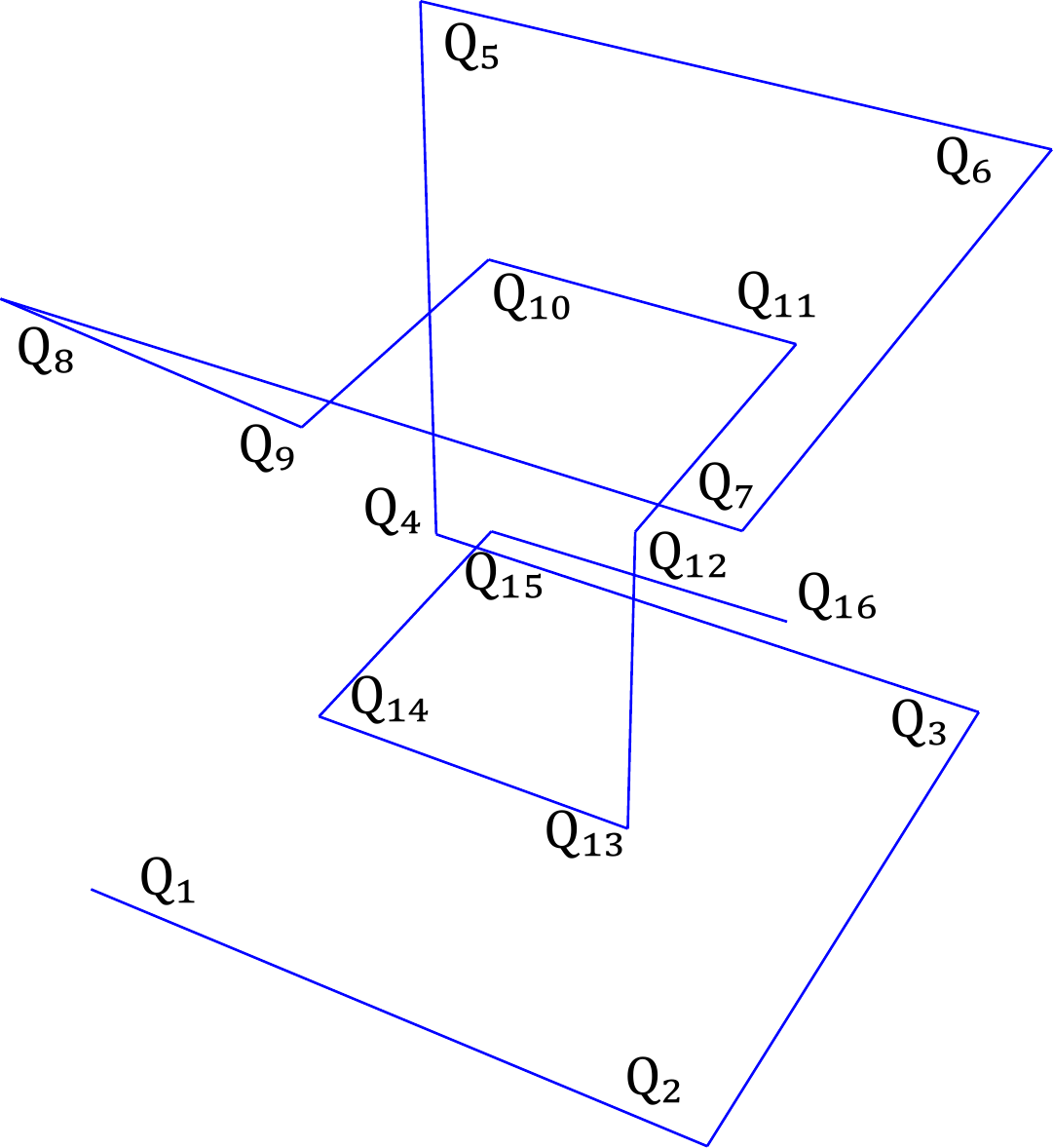}}
    }
    \hspace{3cm}
    \subfloat[Hypercube topology.]{\label{Fig:hypercube_complete}
    \fbox{
    \includegraphics[width=5cm]{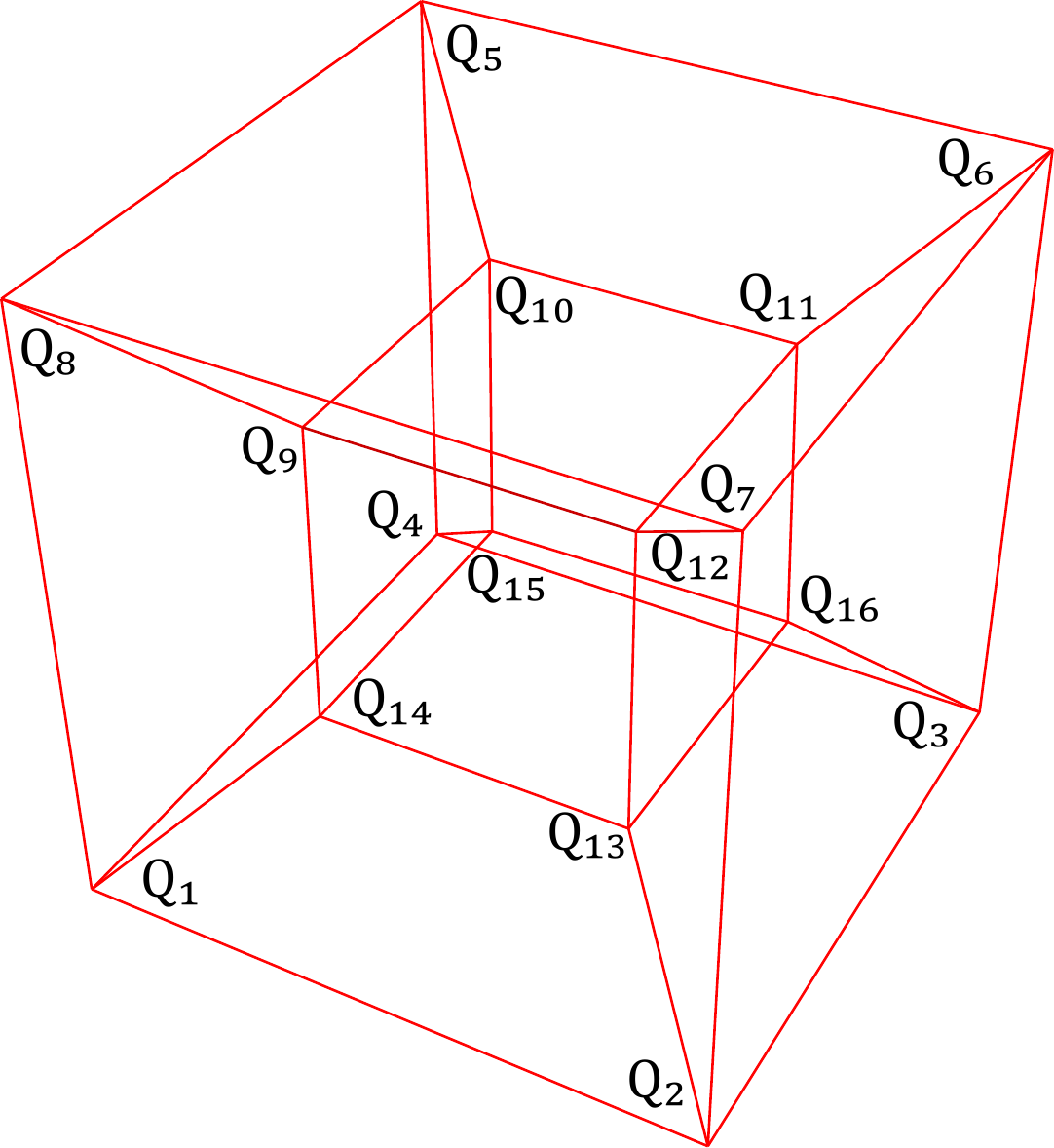}}
    }
    \caption{Comparison of the worst-case scenario linear nearest neighbor topology with the hypercube topology assumed by the compiler proposed by Andr\'es-Mart\'inez \textit{et al.} \cite{AndresMartinez2019}. As shown in Figure~\ref{Fig:hypercube_lnn}, for $n=16$ QPUs the longest path between two QPUs consists of $n-1 = 15$ links, while with the hypercube topology in Figure~\ref{Fig:hypercube_complete} is only about $log_{2}n = 3$ links.}
    \label{Fig:hypercube}
\end{figure*}

More into details, Table~\ref{tab:overhead} reports a sample of the results that have been collected by compiling the circuits with both the \textit{entanglement swapping based} and the \textit{data-qubit swapping based} strategies. Within the table, the first column shows the name of the circuit and the second column shows the number $n$ of logical qubits within the circuit. The third column shows the number of CNOT layers in the original circuit -- i.e., the uncompiled circuit. The fourth and the seventh columns show the theoretical upper bound of the depth of the compiled CNOT layers -- computed in agreement with equations \eqref{eq:4.1} and \eqref{eq:4.3}, respectively -- whereas the fifth and the eighth column shows the depth of the compiled CNOT layers.  For computing the upper bound values and collecting the experimental results, we set the parameters $c_{le}$, $c_{bsm}$ and $c_{cx}$ in equations \eqref{eq:4.2}, \eqref{eq:4.5} and \eqref{eq:4.6} as unit factors, thus obtaining $d_{es} = 3$, $d_{qs} = 9$ and $d'_{qs} = 2$.

Table~\ref{tab:overhead} clearly shows that the upper bounds on the number of layers that result from compiling the layers of remote CNOTs are widely respected and hold, for all the considered examples. Indeed, by comparing the actual depth with the theoretical one, it becomes evident the overestimation of the bounds given in Section~\ref{sec:4.2}. The rationale for this lays in the number of \texttt{CNOT}s in each layer, which are usually significantly lower than what assumed. For instance, let us consider the \textit{GHZ} circuits, where each layer contains a single \texttt{CNOT} and hence the derived bounds -- by assuming $n/2$ \texttt{CNOT}s in each layer -- overestimate the depth. Nevertheless, the discrepancy can be easily fixed by substituting the $n/2$ factor with the actual estimation on the average number of \texttt{CNOT}s in each layer with no loss of generality.

\subsection{Experimental Results for the Worst-Case Topology}
\label{sec:5.3}

\begin{figure*}[ht!]
\centering
\begin{tabular}{ ccc }
    \begin{minipage}{5.4cm}
    	\centering
    	\includegraphics[width=5.4cm]{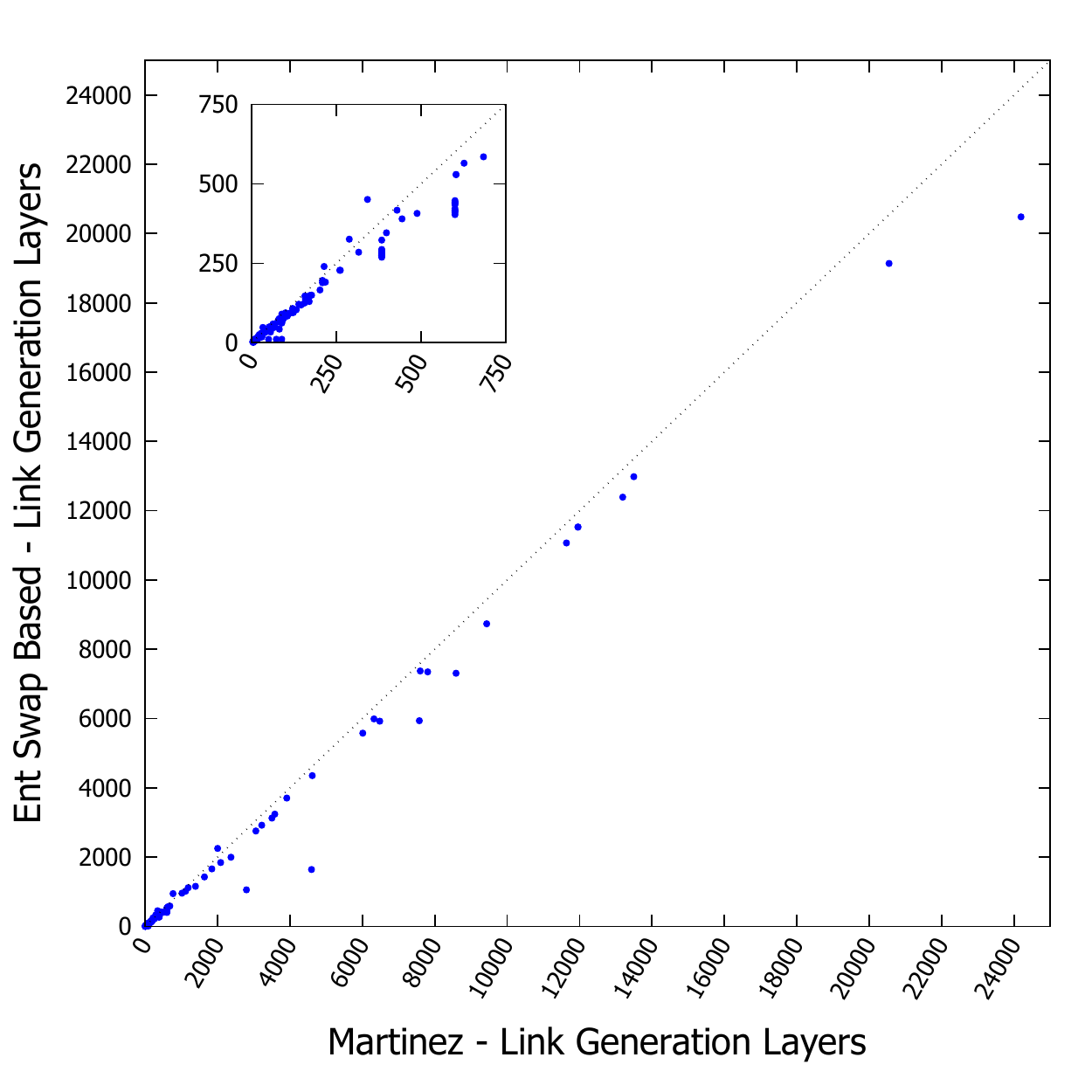}
    	\subcaption{}
        \label{Fig:pure_mart_link}
    \end{minipage}
    &
    \begin{minipage}{5.4cm}
    	\centering
    	\includegraphics[width=5.4cm]{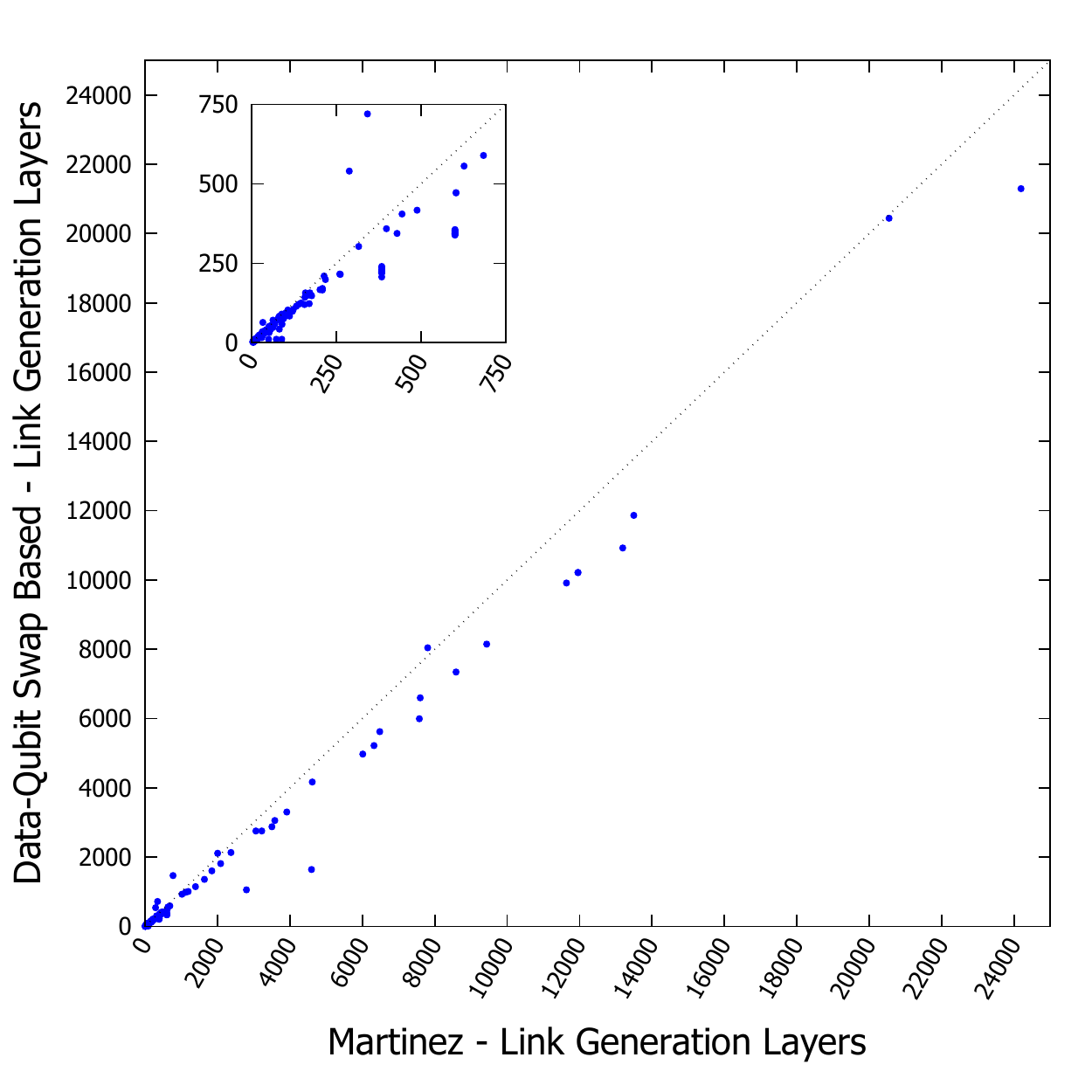}
    	\subcaption{}
        \label{Fig:sort_mart_link}
    \end{minipage}
    &
    \begin{minipage}{5.4cm}
		\centering
		\includegraphics[width=5.4cm]{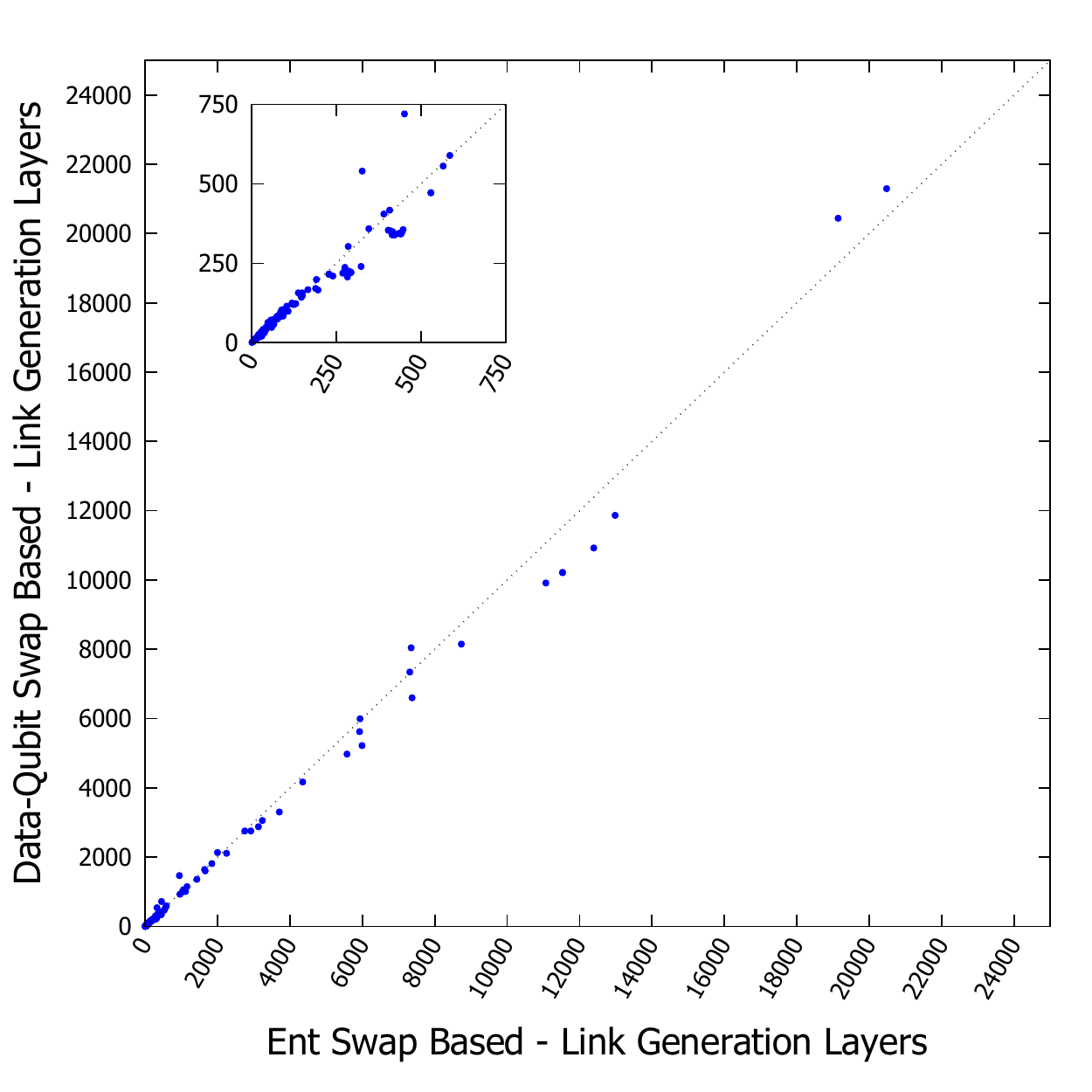}
		\subcaption{}
		\label{Fig:sort_pure_link}
	\end{minipage}
	\end{tabular}
    \caption{Comparing the \textit{entanglement swapping based} and \textit{data-qubit swapping based} strategies of our compiler with Andr\'es-Mart\'inez's compiler~\cite{AndresMartinez2019}, in terms of link entanglement generation layers. Our compiler distributes circuits on the worst-case topology, with only one link between neighboring QPUs (illustrated in Figure~\ref{Fig:07}), while Andr\'es-Mart\'inez's one exploits a more favorable topology, i.e., the hypercube illustrated in Figure~\ref{Fig:hypercube_complete}. Both topologies are characterized by one data qubit per QPU.}
    \label{Fig:link}
\end{figure*}

\begin{figure*}[ht!]
\centering
\begin{tabular}{ ccc }
    \begin{minipage}{5.4cm}
    	\centering
    	\includegraphics[width=5.4cm]{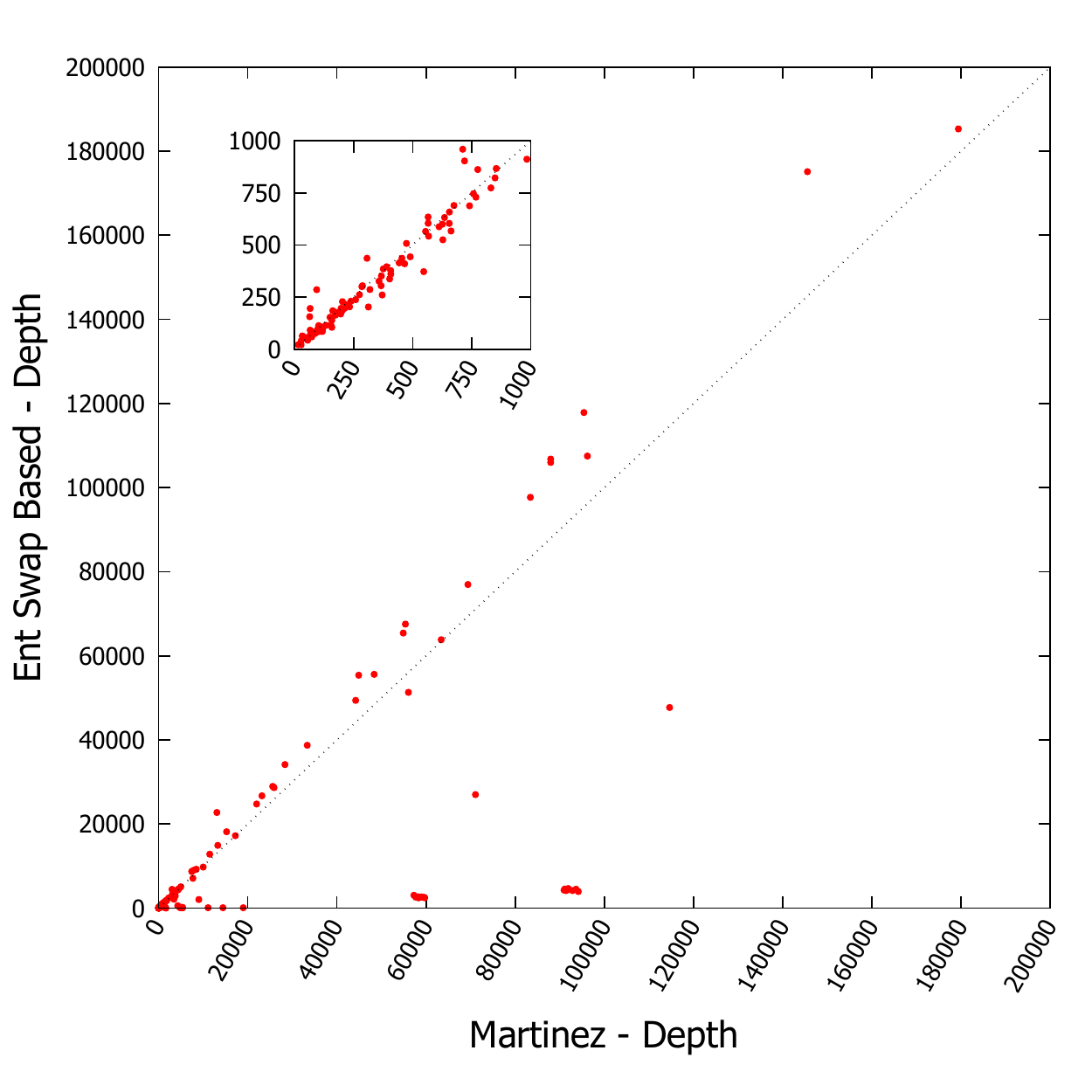}
    	\subcaption{}
        \label{Fig:pure_mart_depth}
    \end{minipage}
    &
    \begin{minipage}{5.5cm}
    	\centering
    	\includegraphics[width=5.4cm]{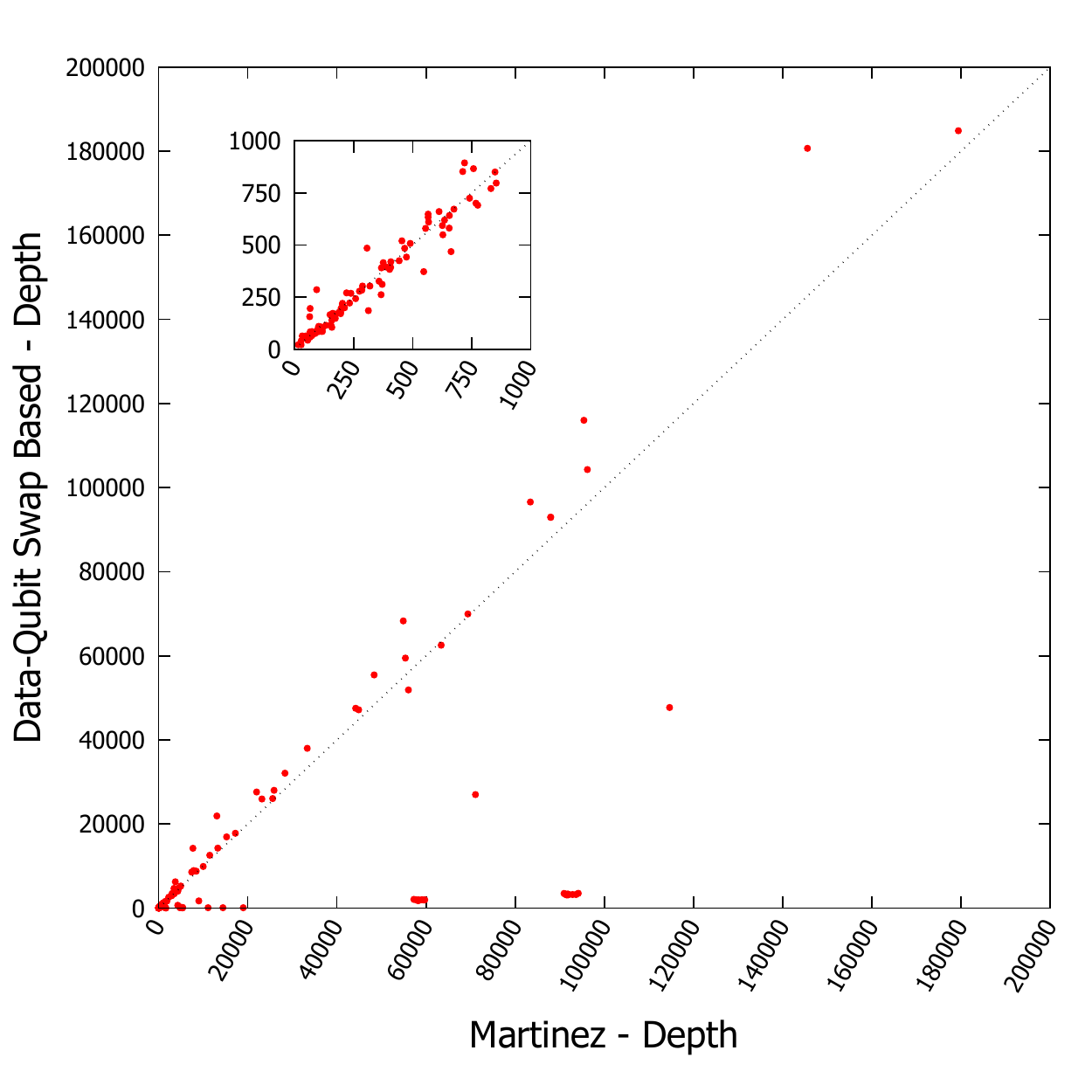}
    	\subcaption{}
        \label{Fig:sort_mart_depth}
    \end{minipage}
    &
    \begin{minipage}{5.4cm}
		\centering
		\includegraphics[width=5.4cm]{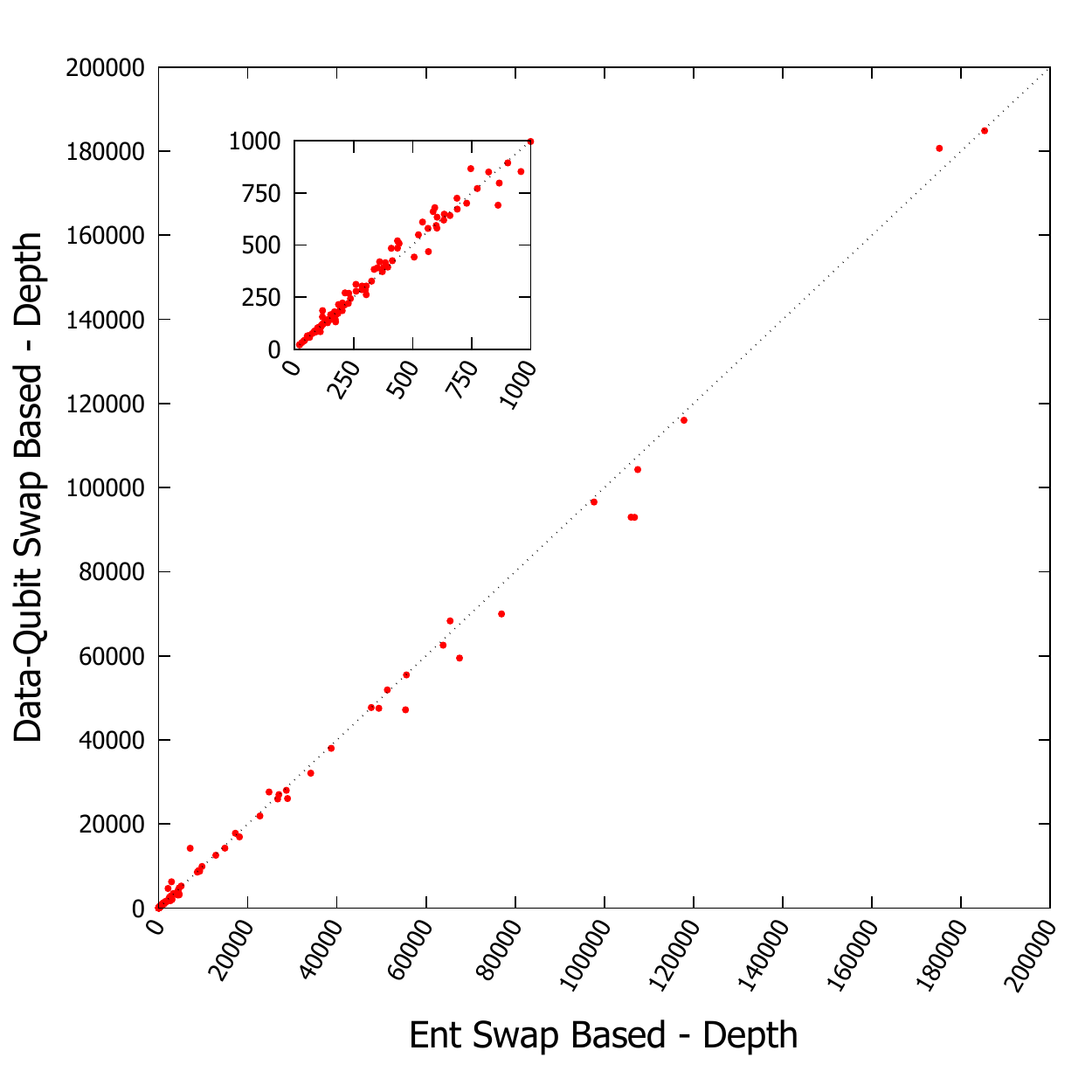}
		\subcaption{}
		\label{Fig:sort_pure_depth}
	\end{minipage}
	\end{tabular}
	\caption{Comparing the \textit{entanglement swapping based} and \textit{data-qubit swapping based} strategies of our compiler with Andr\'es-Mart\'inez's compiler~\cite{AndresMartinez2019}, in terms of circuit depth. Our compiler distributes circuits on the worst-case topology, with only one link between neighboring QPUs (illustrated in Figure~\ref{Fig:07}), while Andr\'es-Mart\'inez's one exploits a more favorable topology, i.e., the hypercube illustrated in Figure~\ref{Fig:hypercube_complete}. Both topologies are characterized by one data qubit per QPU.}
    \label{Fig:depth}
\end{figure*}

\begin{figure*}[ht!]
\centering
\begin{tabular}{ ccc }
    \begin{minipage}{5.4cm}
    	\centering
    	\includegraphics[width=5.4cm]{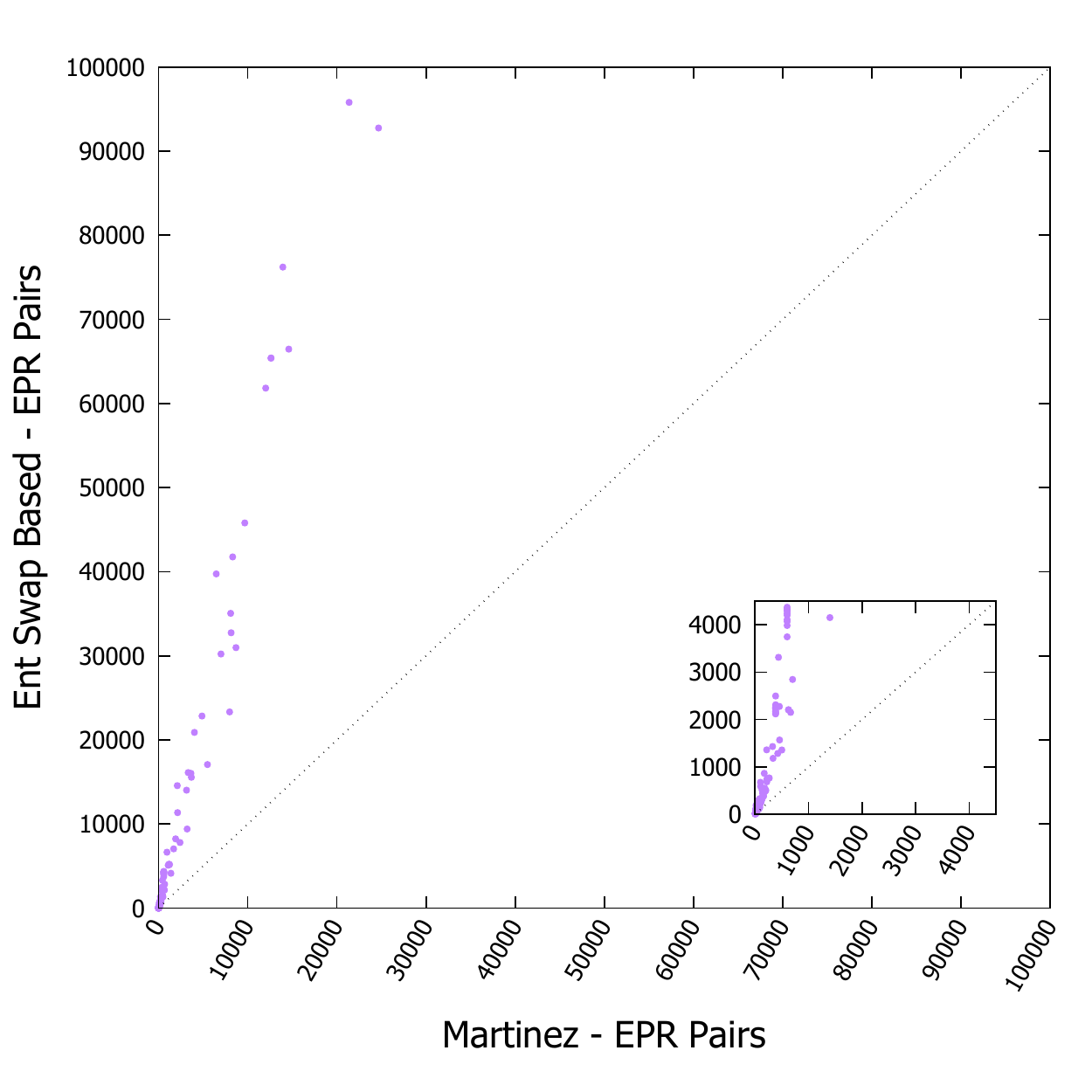}
    	\subcaption{}
        \label{Fig:pure_mart_epr}
    \end{minipage}
    &
    \begin{minipage}{5.4cm}
    	\centering
    	\includegraphics[width=5.4cm]{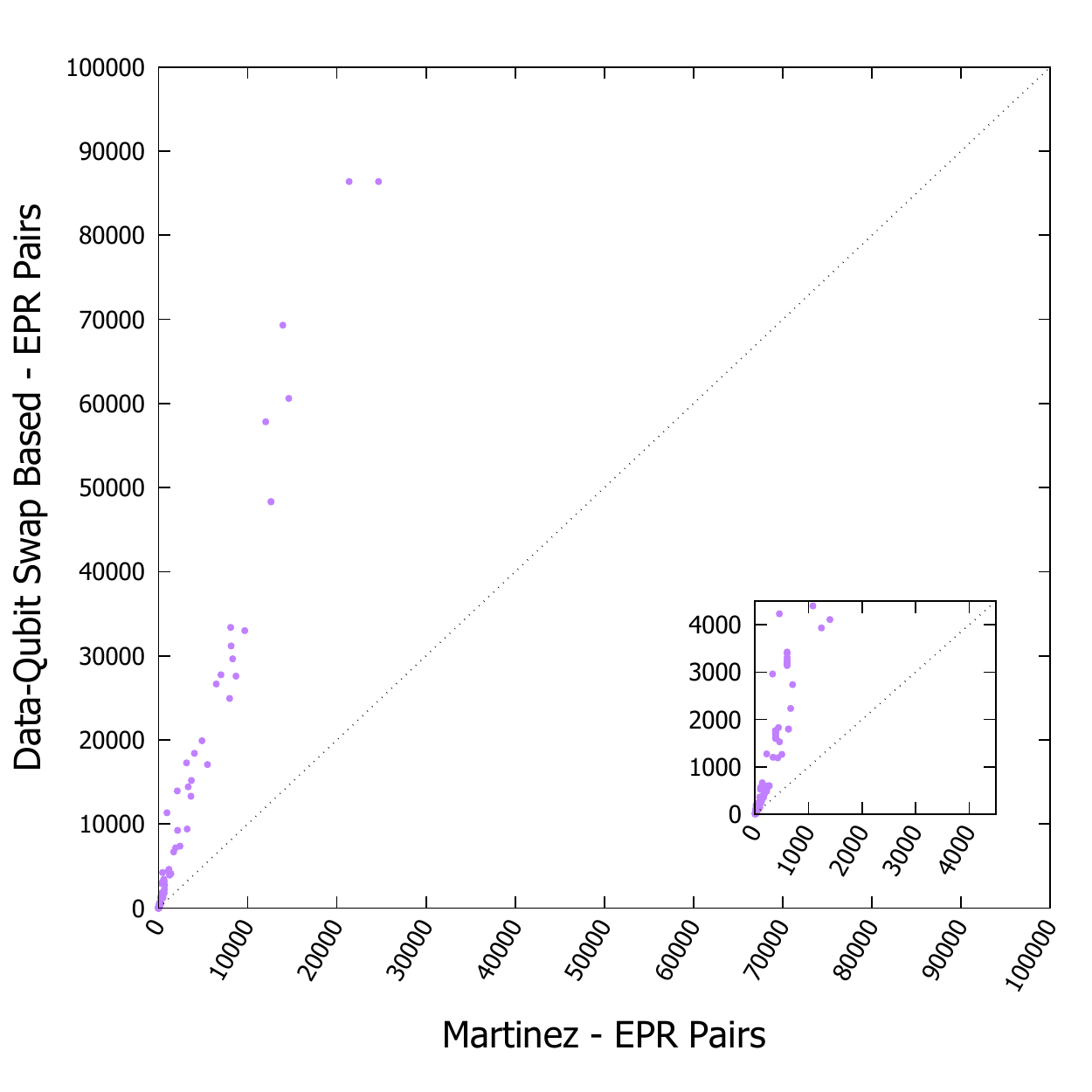}
    	\subcaption{}
        \label{Fig:sort_mart_epr}
    \end{minipage}
    &
    \begin{minipage}{5.4cm}
		\centering
		\includegraphics[width=5.4cm]{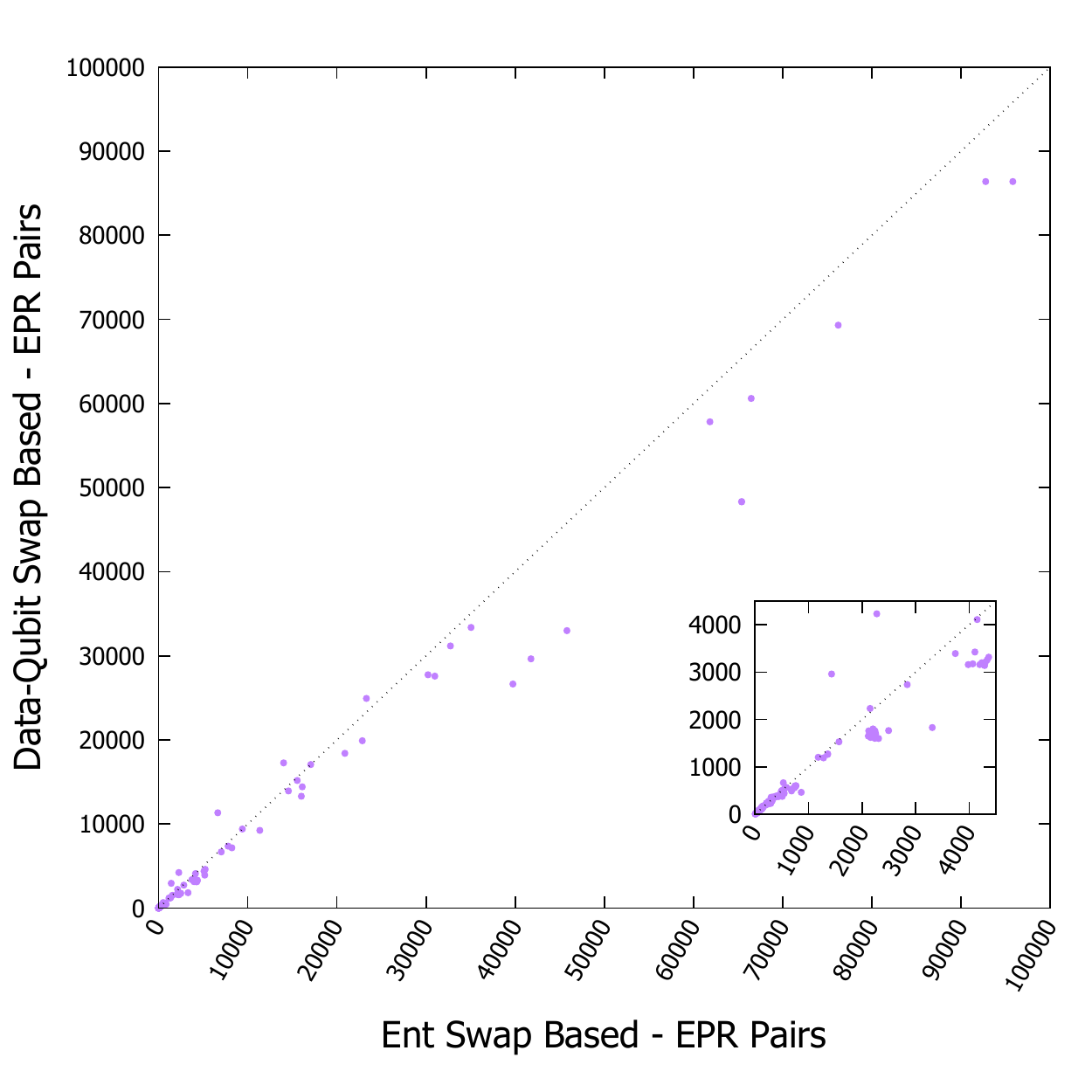}
		\subcaption{}
		\label{Fig:sort_pure_epr}
	\end{minipage}
	\end{tabular}
    \caption{Comparing the \textit{entanglement swapping based} and \textit{data-qubit swapping based} strategies of our compiler with Andr\'es-Mart\'inez's compiler~\cite{AndresMartinez2019}, in terms of consumed EPR pairs. Our compiler distributes circuits on the worst-case topology, with only one link between neighboring QPUs (illustrated in Figure~\ref{Fig:07}), while Andr\'es-Mart\'inez's one exploits a more favorable topology, i.e., the hypercube illustrated in Figure~\ref{Fig:hypercube_complete}. Both topologies are characterized by one data qubit per QPU.}
    \label{Fig:epr}
\end{figure*}

We compared our compiler with the one proposed by Andr\'es-Mart\'inez \textit{et al.} \cite{AndresMartinez2019}, in respect of which we were able to set each QPU memory to one data qubit but we could not impose any limitation over the number of communication qubits per QPU nor the topology of the quantum network, which is always assumed to be an hypercube. Such a topology is shown in Figure~\ref{Fig:hypercube_complete}, where one can clearly see the connectivity disparity, compared to the worst-case topology illustrated in Figure~\ref{Fig:hypercube_lnn}. In Figures~\ref{Fig:link}-\ref{Fig:epr}, the results of the comparative evaluation are plotted. For a better readability of the figures, we omit data related to a 20 qubit random circuit that presents values far greater than the rest of the data set, for all compiling strategies including the state of the art (such data have been included in Table~\ref{tab:overhead}).

Figure~\ref{Fig:link} shows the number of Link Generation Layers, i.e., the number of layers in the distributed circuit that comprise only Bell state generation and distribution between QPUs. It is clear that our compiler requires less layers of link generation for almost every tested circuit. Having fewer layers of link generation reduces the time that data qubits have to spend idle, i.e., possibly affected by decoherence, while waiting to be able to perform remote operations. Figure~\ref{Fig:sort_pure_link} also shows that choosing the \textit{data-qubit swapping based strategy} against the \textit{entanglement swapping based strategy} is generally the best choice.

Regarding the depth of the distributed circuits, illustrated in Figure~\ref{Fig:depth}, we can see that for some circuits our compiler clearly outperforms Andr\'es-Mart\'inez's one. Figure~\ref{Fig:sort_pure_depth} confirms that the \textit{data-qubit swapping based strategy} is better than the \textit{entanglement swapping based} one.

\begin{figure*}[ht!]
\centering
	\begin{tabular}{ ccc }
	\begin{minipage}{5.4cm}
    	\centering
    	\includegraphics[width=5.4cm]{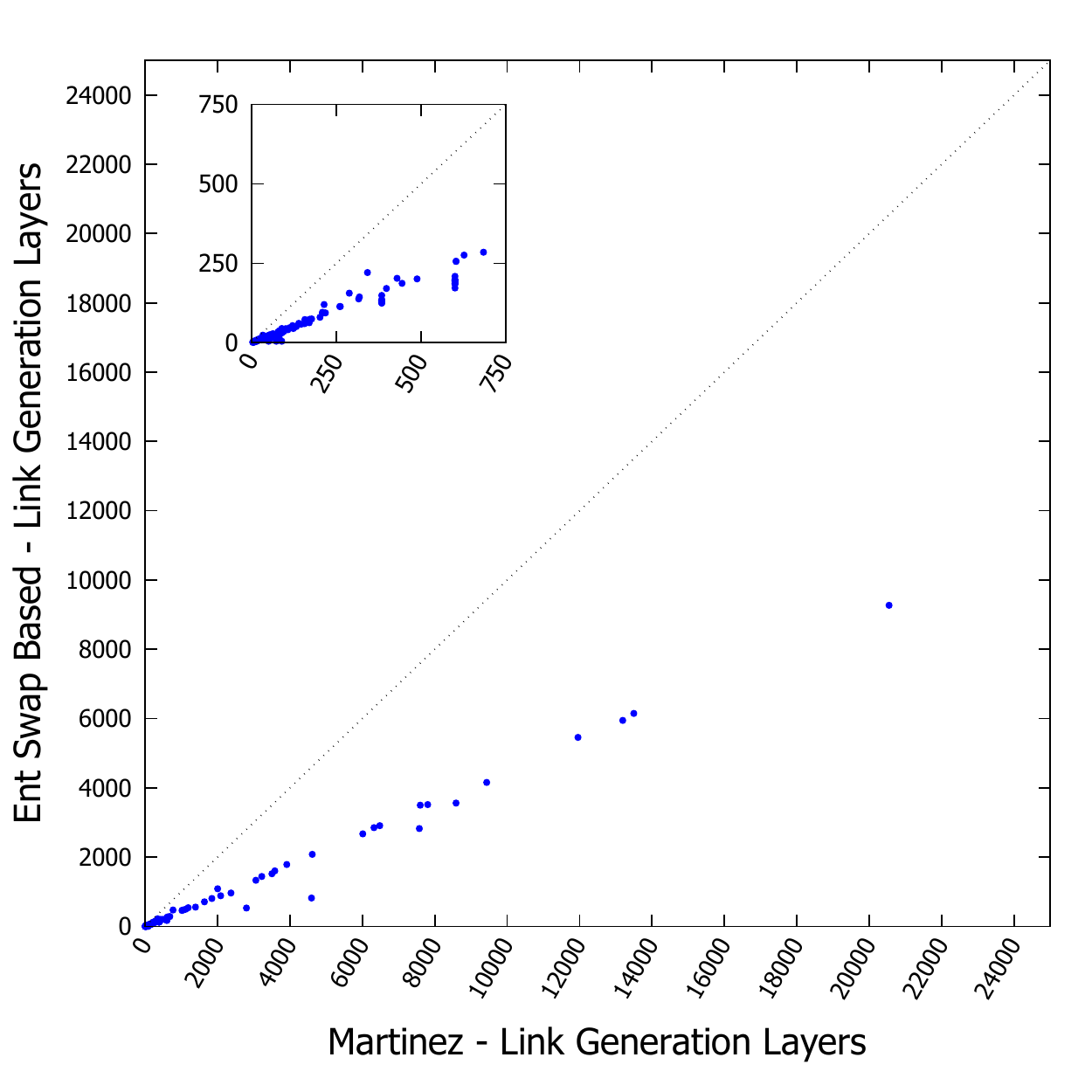}
    	\subcaption{}
        \label{Fig:pure_mart_link_2ebits}
    \end{minipage}
    &
    \begin{minipage}{5.4cm}
    	\centering
    	\includegraphics[width=5.4cm]{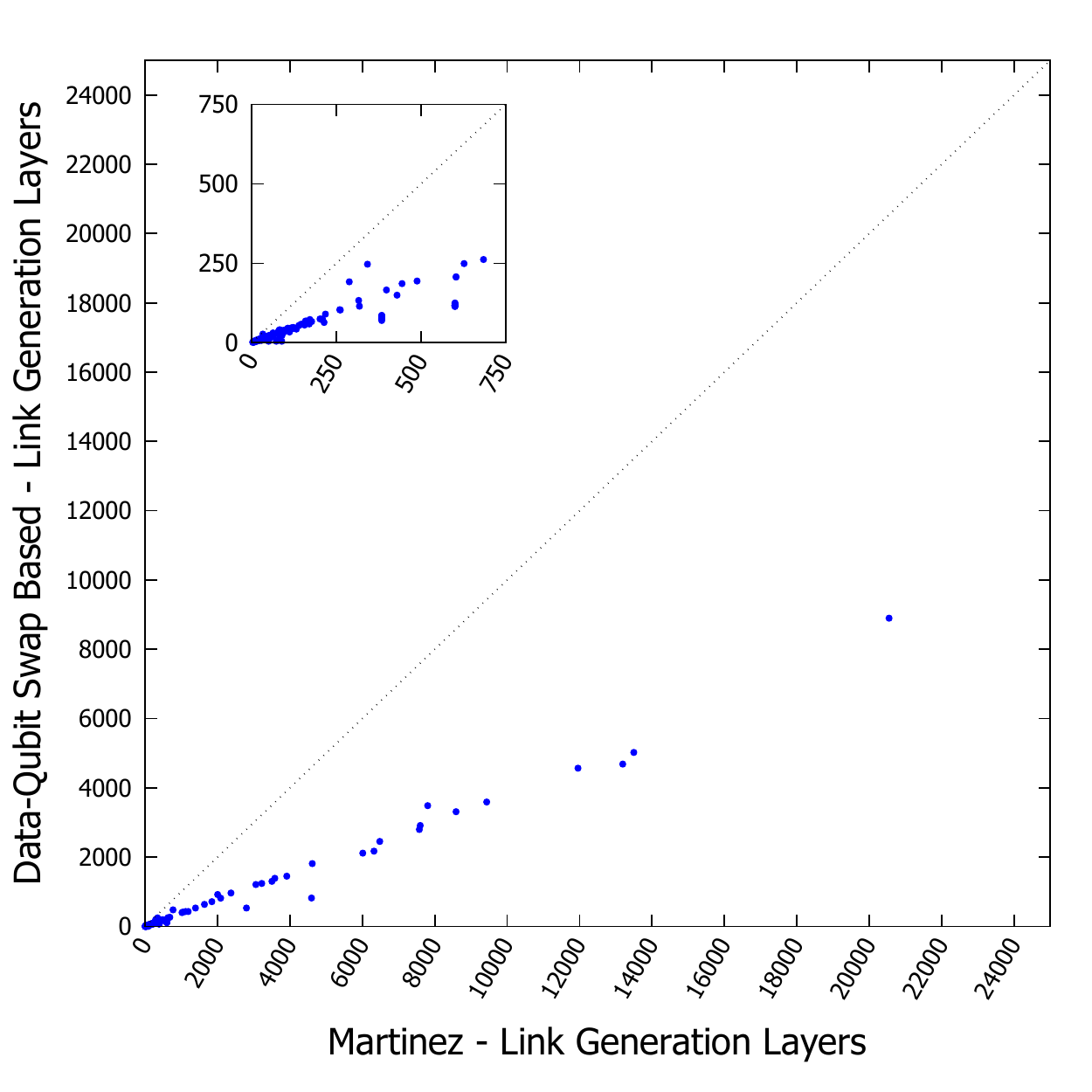}
    	\subcaption{}
        \label{Fig:sort_mart_link_2ebits}
    \end{minipage}
    &
    \begin{minipage}{5.4cm}
		\centering
		\includegraphics[width=5.4cm]{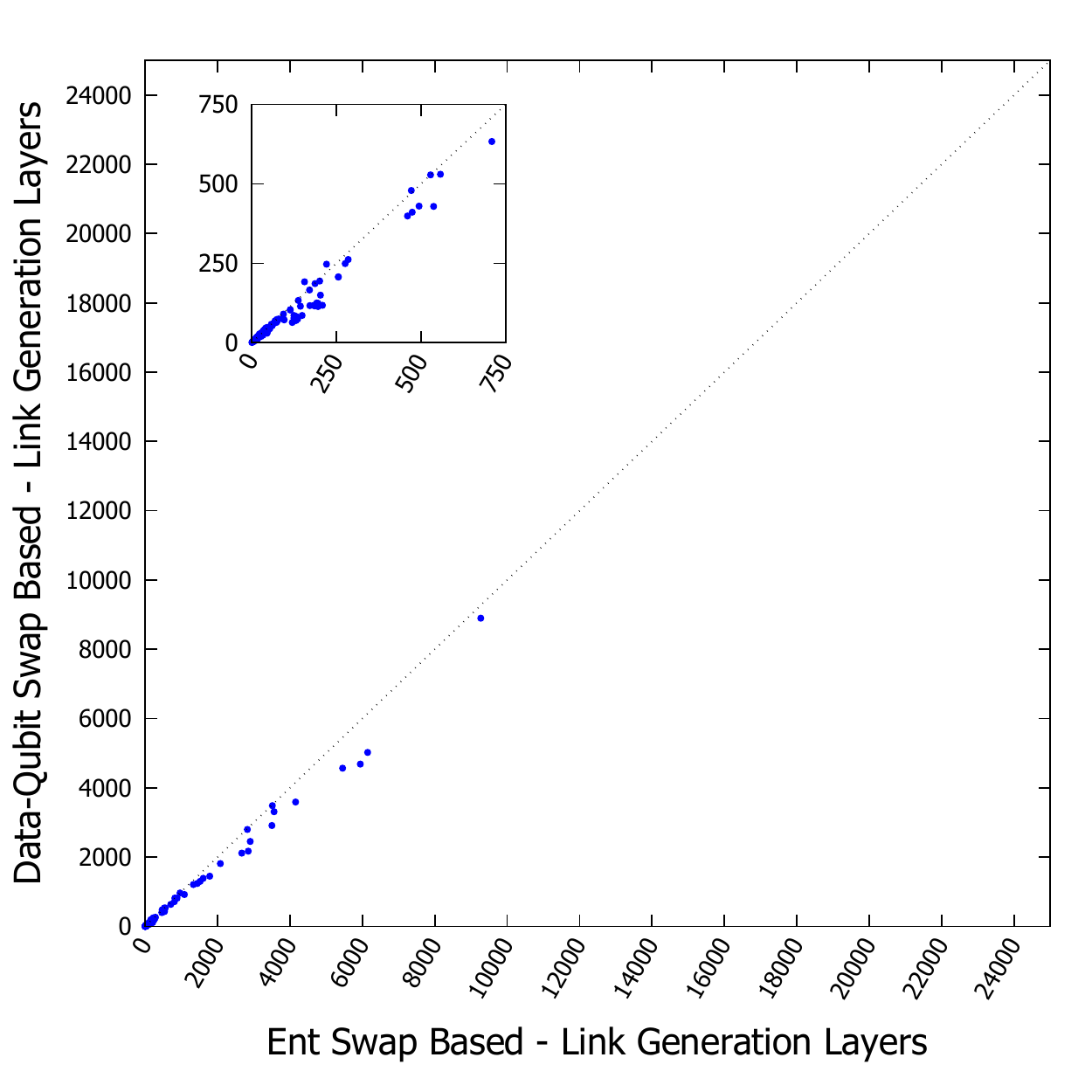}
		\subcaption{}
		\label{Fig:sort_pure_link_2ebits}
	\end{minipage}
	\end{tabular}
    \caption{Comparing the \textit{entanglement swapping based} and \textit{data-qubit swapping based} strategies of our compiler with Andr\'es-Mart\'inez's compiler~\cite{AndresMartinez2019} over link entanglement generation layers. Our compiler distributed circuits on the topology illustrated in Figure~\ref{Fig:12}, with two links between neighboring QPUs(illustrated in Figure~\ref{Fig:07}), while Andr\'es-Mart\'inez's one exploits a more favorable topology, i.e., the hypercube illustrated in Figure~\ref{Fig:hypercube_complete}. Both topologies are characterized by one data qubit per QPU.}
    \label{Fig:link_2ebits}
\end{figure*}

With respect to the number of generated Bell states, depicted in Figure~\ref{Fig:epr}, it can be observed that our compiler consumes a fair amount of Bell states compared to Andr\'es-Mart\'inez's compiler. This was expected and it is mostly due to the fact that Andr\'es-Mart\'inez's compiler benefits from a hypercube network topology, as showed in Figure~\ref{Fig:hypercube_complete}. Using such a topology means that, in most cases, a link between two QPUs can be directly generated with just one Bell state, which is in direct contrast with the worst-case topology that we used, depicted in Figure~\ref{Fig:hypercube_lnn}. In our linear topology, to generate a link between two non neighboring QPUs we need to perform entanglement swapping, generating and consuming Bell states shared by all the others QPUs in between. Nevertheless, the time needed to generate one Bell state should be the same as to generate $n$ Bell states in parallel, and with Figure~\ref{Fig:link} we already showed that our compiler usually needs fewer layers of link generation.

It is worthwhile to note that with network topologies different from the considered one -- namely, the worst-case topology  where each \texttt{CNOT} must be mapped into a remote \texttt{CNOT} -- new optimization challenges arise. As instance, whenever multiple data qubits are available at each (or some nodes), only a subset of \texttt{CNOT}s must be mapped into remote operations. Hence, the compiler should be able to optimize choices such as which sub-circuit should be mapped to which node or which \texttt{CNOT} should be performed via communication qubits. Clearly, the optimal strategies -- as well as the metrics to measure the optimality of a strategy -- represents interesting open problems.

\begin{figure*}[ht!]
\centering
	\begin{tabular}{ ccc }
	\begin{minipage}{5.4cm}
    	\centering
    	\includegraphics[width=5.4cm]{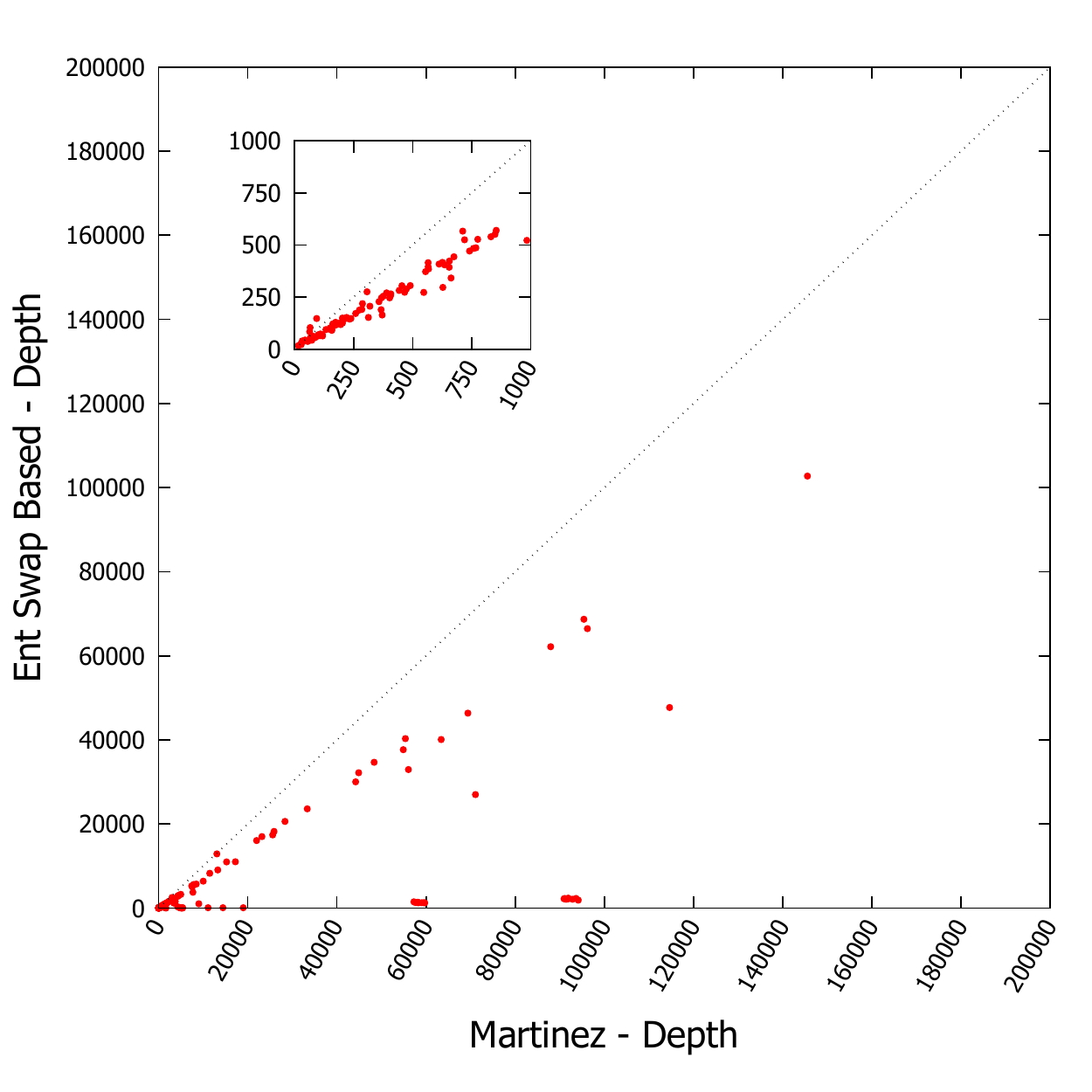}
    	\subcaption{}
        \label{Fig:pure_mart_depth_2ebits}
    \end{minipage}
    &
    \begin{minipage}{5.4cm}
    	\centering
    	\includegraphics[width=5.4cm]{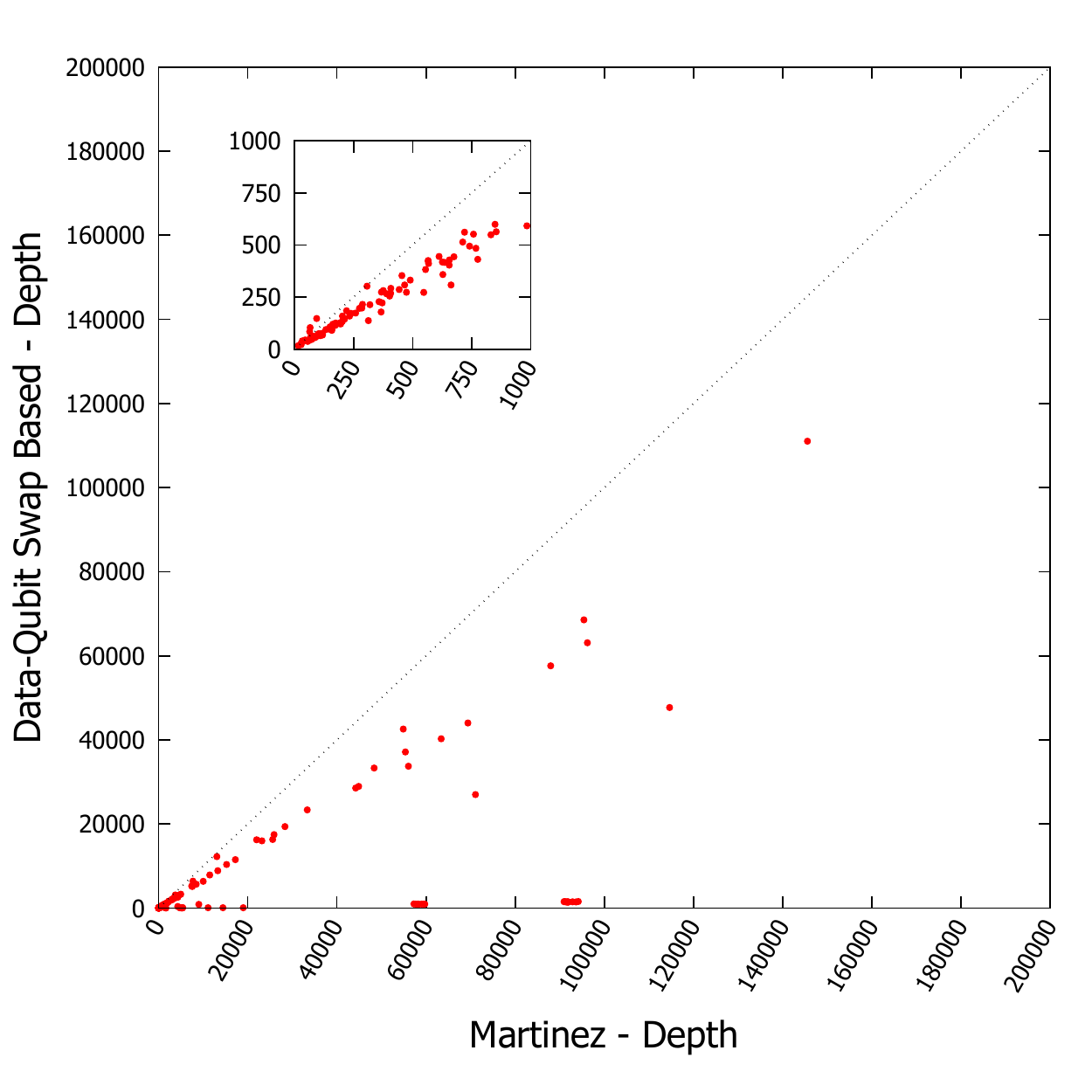}
    	\subcaption{}
        \label{Fig:sort_mart_depth_2ebits}
    \end{minipage}
    &
    \begin{minipage}{5.4cm}
		\centering
		\includegraphics[width=5.4cm]{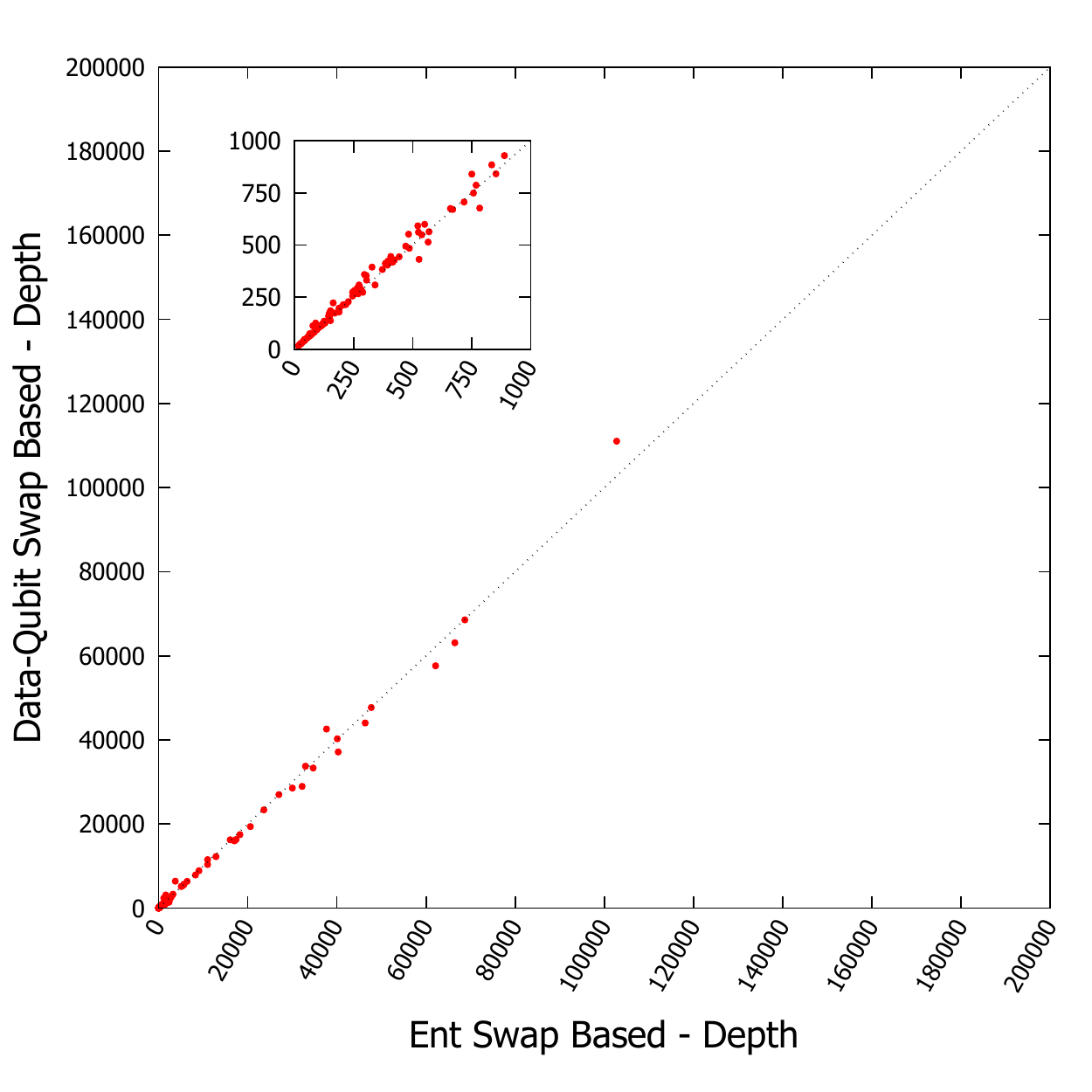}
		\subcaption{}
		\label{Fig:sort_pure_depth_2ebits}
	\end{minipage}
	\end{tabular}
    \caption{Comparing the \textit{entanglement swapping based} and \textit{data-qubit swapping based} strategies of our compiler with Andr\'es-Mart\'inez's compiler~\cite{AndresMartinez2019}, in terms of circuits' depth. Our compiler distributed circuits on the topology illustrated in Figure~\ref{Fig:12}, with two links between neighboring QPUs (illustrated in Figure~\ref{Fig:07}), while Andr\'es-Mart\'inez's one exploits a more favorable topology, i.e., the hypercube illustrated in Figure~\ref{Fig:hypercube_complete}. Both topologies are characterized by one data qubit per QPU.}
    \label{Fig:depth_2ebits}
\end{figure*}

\begin{figure*}[ht!]
\centering
	\begin{tabular}{ ccc }
	\begin{minipage}{5.4cm}
    	\centering
    	\includegraphics[width=5.4cm]{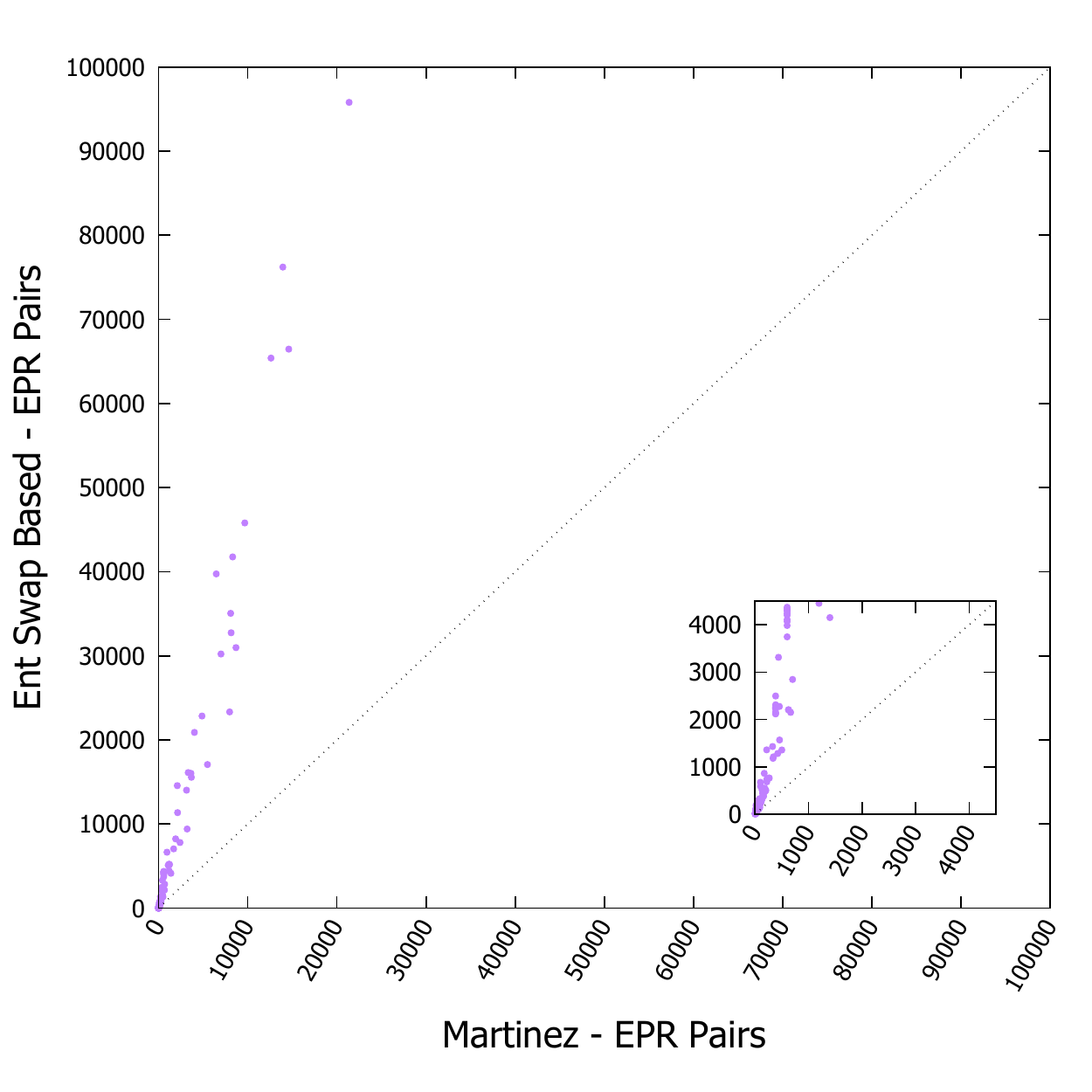}
    	\subcaption{}
        \label{Fig:pure_mart_epr_2ebits}
    \end{minipage}
    &
    \begin{minipage}{5.4cm}
    	\centering
    	\includegraphics[width=5.4cm]{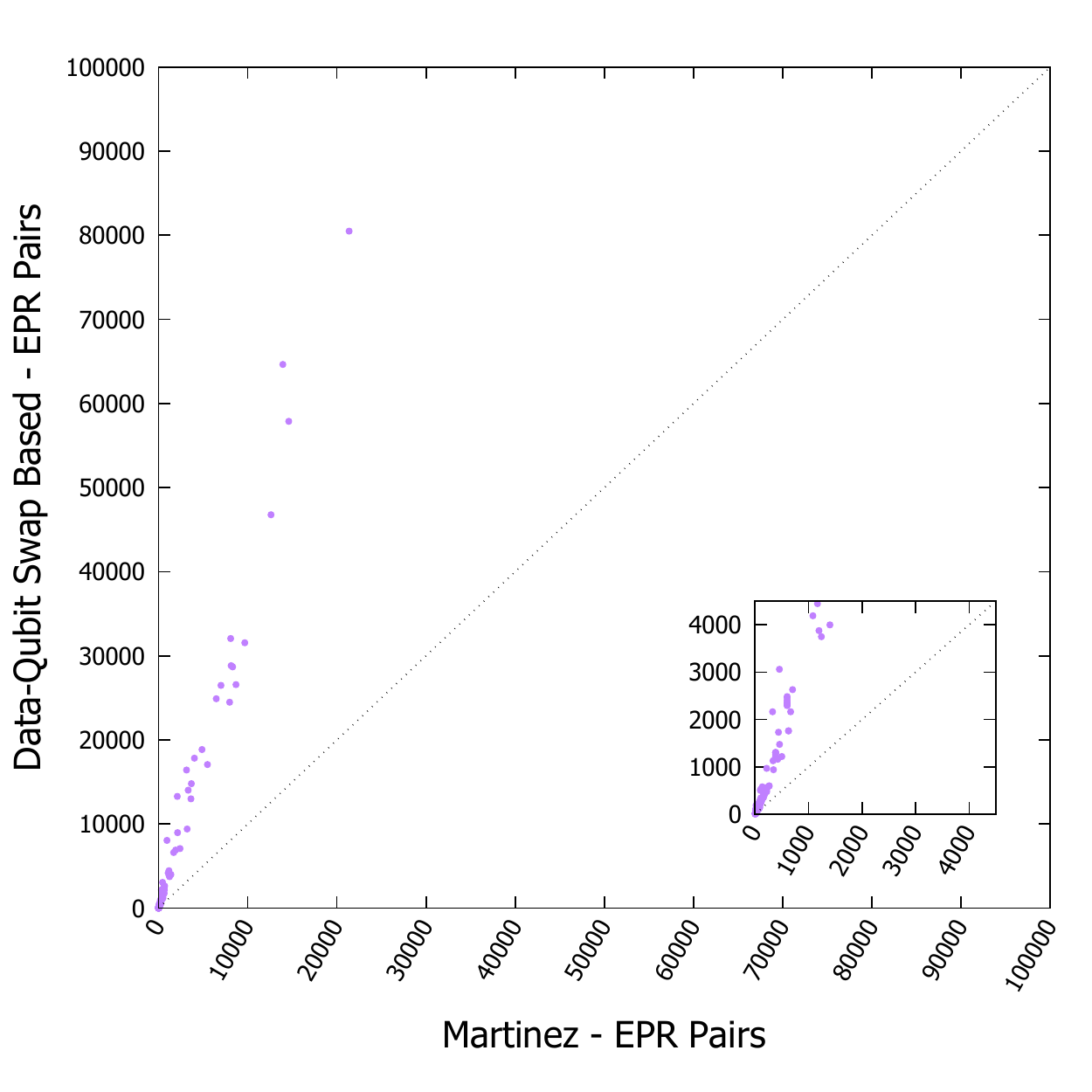}
    	\subcaption{}
        \label{Fig:sort_mart_epr_2ebits}
    \end{minipage}
    &
    \begin{minipage}{5.4cm}
		\centering
		\includegraphics[width=5.4cm]{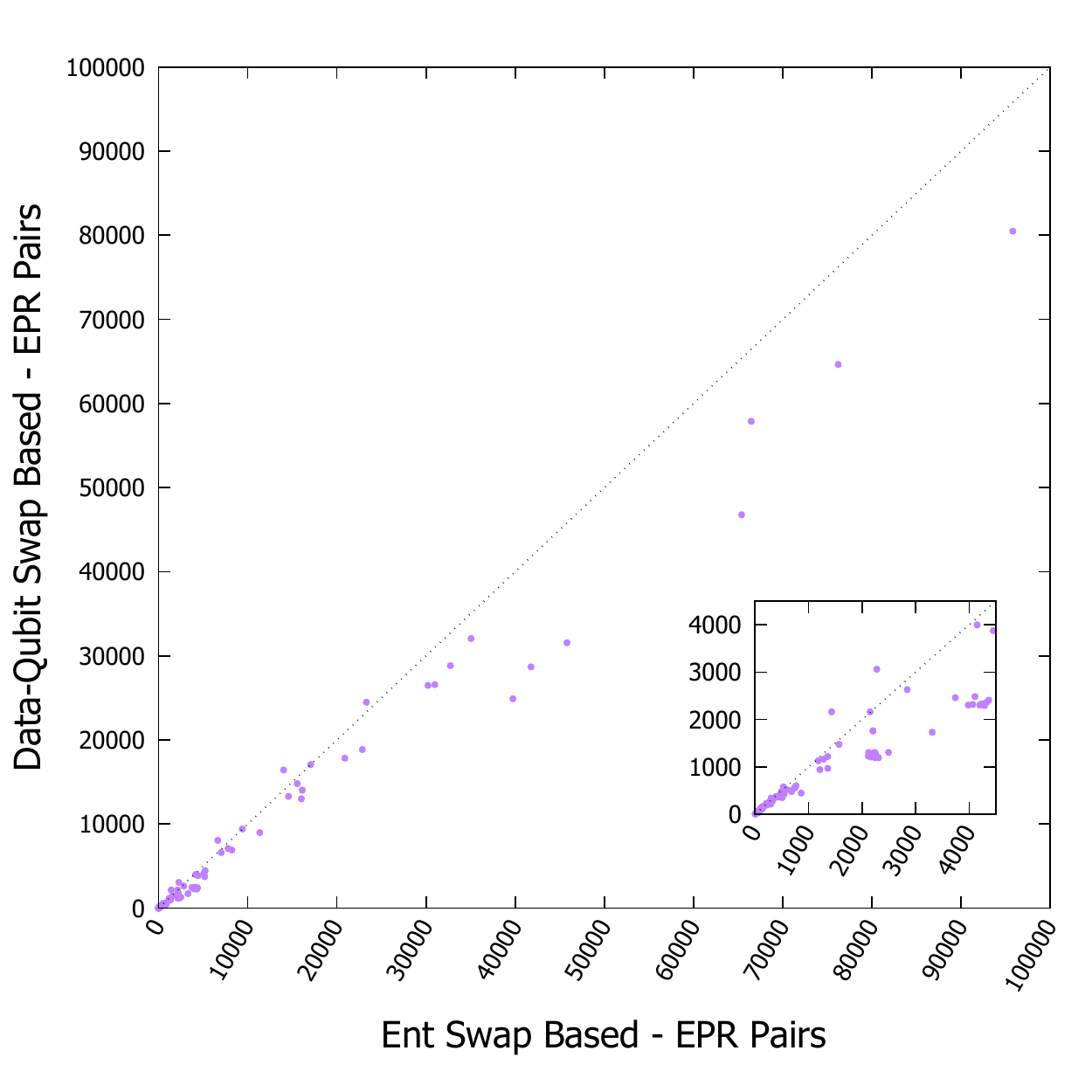}
		\subcaption{}
		\label{Fig:sort_pure_epr_2ebits}
	\end{minipage}
	\end{tabular}
    \caption{Comparing the \textit{entanglement swapping based} and \textit{data-qubit swapping based} strategies of our compiler with Andr\'es-Mart\'inez's compiler~\cite{AndresMartinez2019} over consumed EPR pairs. Our compiler distributed circuits on the topology illustrated in Figure~\ref{Fig:12}, with two links between neighboring QPUs (illustrated in Figure~\ref{Fig:07}), while Andr\'es-Mart\'inez's one exploits a more favorable topology, i.e., the hypercube illustrated in Figure~\ref{Fig:hypercube_complete}. Both topologies are characterized by one data qubit per QPU.}
    \label{Fig:epr_2ebits}
\end{figure*}

\subsection{Additional Experimental Results}
\label{sec:5.4}

To show that the proposed compiling strategies can be applied to more complex topologies, we tested them on a slight variation of the worst-case scenario. This new topology is depicted in Figure~\ref{Fig:12}, where we doubled the number of EPR pairs per QPU, hence we doubled the number of links between neighbor QPUs.
As described in Section~\ref{sec:5.1}, this setting enables our compiler to perform data-qubit swapping in a more efficient way and also greatly reduces the number of layers dedicated to the link entanglement generation, as we can generate and use double the number of links in parallel. Such a performance improvement is clearly shown in Figure~\ref{Fig:link_2ebits}.

The same considerations apply to the depth of the compiled circuits, illustrated in Figure~\ref{Fig:depth_2ebits}, where we can observe an appreciable improvement against the worst-case scenario and the state of the art compiler. As for the number of generated EPR pairs, aside from an imperceptible difference with the worst-case scenario, the advantage of an hypercube topology is still evident.

\section{Conclusion}
\label{sec:6}
In this paper, we have discussed the main challenges arising with compiler design for distributed quantum computing. Then, we analytically derived an upper bound of the overhead induced by quantum compilation for distributed quantum computing. The derived bound accounts for the overhead induced by the underlying computing architecture as well as the additional overhead induced by the sub-optimal quantum compiler. To this aim, we designed a quantum compiler with three key features: i) general-purpose,  namely,  requiring  no  particular  assumptions on the quantum circuits to be compiled, ii) efficient, namely, exhibiting a polynomial-time computational complexity so that it can successfully compile medium-to-large circuits of practical value, and iii) effective,  being  the  total  circuit  depth  overhead  induced by the quantum circuit compilation always upper-bounded by a factor that grows linearly with the number of logical qubits of the original quantum circuit. We validated the theoretical upper bound against an extensive set of medium-size quantum circuits of practical interest, and we confirmed the validity of the compiler design through an extensive performance analysis.

\section*{Acknowledgments}
The authors would like to thank Pablo Andr\'es-Mart\'inez for answering questions about his related work.


\newpage

\section{Appendix A}
\label{app:1}
Here we present a pseudocode description of the whole compilation process illustrated in Figure~\ref{Fig:compiler_workflow} and summarized by Algorithm~\ref{alg:compile}. The front layer is iteratively updated with Algorithm~\ref{alg:update_front} and compiled by Algorithm~\ref{alg:compile_front}. This is done until no more front layers can be computed, meaning that all gates have been mapped and the compilation process has finished.

\begin{algorithm}
    \caption{\textsc{Compile}\newline
    \footnotesize
    \textbf{Input}: the $circuit$ to be compiled\newline
    \textbf{Output}: the compiled circuit}
    \label{alg:compile}
    \begin{algorithmic}
    \footnotesize
    \State $\mathcal{G} \gets$ all gates from $circuit$
    \State $\mathcal{E} \gets \emptyset$ \Comment{executed gates}
    \State create $new\_circuit$ as an empty circuit
    \While{$\mathcal{G} \neq \emptyset$}
        \State $\mathcal{G}, \mathcal{F}, \mathcal{E} \gets$ \Call{UpdateFrontLayer}{$\mathcal{G}, \mathcal{E}$} \Comment{$\mathcal{F}$ is the front layer}
        \If{$\mathcal{F} \neq \emptyset$}
            \State $\mathcal{E} \gets$ \textsc{CompileFrontLayer($\mathcal{F}, \mathcal{E}$)}
        \EndIf
    \EndWhile
    \ForAll{$gate \in \mathcal{E}$}:
        \State add $gate$ to $new\_circuit$
    \EndFor
    \State \textbf{return} $new\_circuit$
    \end{algorithmic}
\end{algorithm}

\begin{algorithm}
    \caption{\textsc{UpdateFrontLayer}\newline
    \footnotesize
    \textbf{Input}: $\mathcal{G}$ gates to be executed, $\mathcal{E}$ executed gates until now\newline
    \textbf{Output}: updated $\mathcal{G}$, updated $\mathcal{E}$, new front layer $\mathcal{F}$}
    \label{alg:update_front}
    \begin{algorithmic}
    \footnotesize
    \State $\mathcal{A} \gets \emptyset$ \Comment{allocated qubits}
        \State $\mathcal{F} \gets \emptyset$ \Comment{new front layer}
        \State $\mathcal{R} \gets \emptyset$ \Comment{gates to remove}
        \State $width \gets $ total number of data qubits available
        \ForAll{$gate \in \mathcal{G}$}
            \If{$|\mathcal{A}| = width$}
                \State \textbf{break}
            \EndIf
            \If{$\mathcal{A} \cap gate.qubits = \emptyset$}
                \If{$gate$ is a one-qubit gate}
                    \State add $gate$ to $\mathcal{E}$
                \Else
                    \State add $gate$ to $\mathcal{F}$
                    \State add $gate$ to $\mathcal{E}$
                    \State add $gate.qubits$ to $\mathcal{A}$
                \EndIf
                \State add $gate$ to $\mathcal{R}$
            \Else:
                \State add $gate.qubits$ to $\mathcal{A}$
            \EndIf
        \EndFor
        \ForAll{$gate \in \mathcal{R}$}
            \State remove $gate$ from $\mathcal{G}$
        \EndFor
        \State \textbf{return} $\mathcal{G}, \mathcal{F}, \mathcal{E}$
    \end{algorithmic}
\end{algorithm}

\newpage

\begin{small}

\begin{wrapfigure}{l}{25mm} 
    \includegraphics[width=1in,height=1.25in,clip,keepaspectratio]{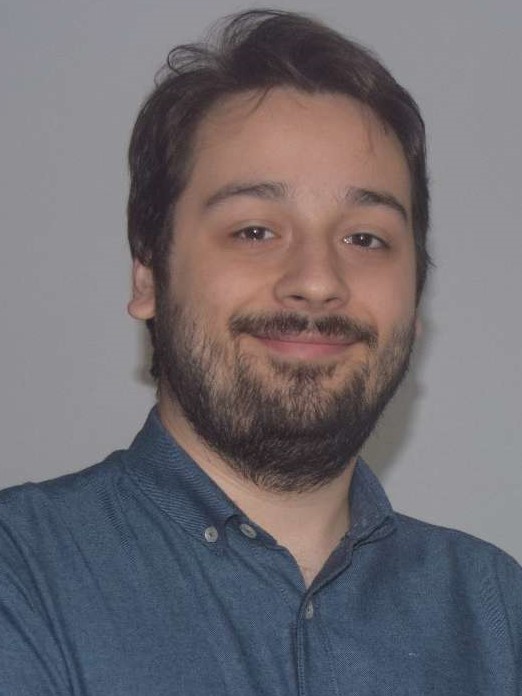}
  \end{wrapfigure}\par
  \textbf{Davide Ferrari} (GS'20) received the BSc degree in Computer Engineering from the Polytechnic of Milan, Italy, in 2016 and the MSc degree in Computer Engineering from the University of Parma, Italy, in 2019. Right after, he has been a research scholar at Future Technology Lab of the University of Parma, working on the design of efficient algorithms for quantum compiling. 
Currently, he is a PhD student of the Department of Engineering and Architecture of the University of Parma. He is involved in the Quantum Information Science (QIS) research initiative at the University of Parma, where he is a member of the Quantum Software research unit. In 2020, he won the 'IBM Quantum Awards Circuit Optimization Developer Challenge'. His research focuses on efficient quantum compiling for quantum simulations and quantum machine learning.\par

\vspace{0.2cm}

\begin{wrapfigure}{l}{25mm} 
    \includegraphics[width=1in,height=1.25in,clip,keepaspectratio]{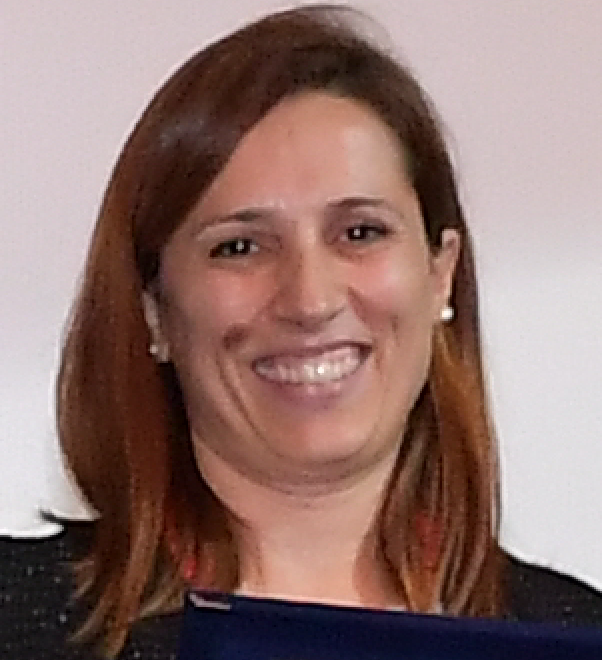}
  \end{wrapfigure}\par
  \textbf{Angela Sara Cacciapuoti} (M'10, SM'16) is a faculty at the University of Naples Federico II, Italy. In 2009, she received the Ph.D. degree in Electronic and Telecommunications Engineering, and in 2005 a `Laurea' (integrated BS/MS) summa cum laude in Telecommunications Engineering, both from the University of Naples Federico II. She was a visiting researcher at Georgia Institute of Technology (USA) and at the Universitat Politecnica de Catalunya (Spain). Since July 2018, she held the national habilitation as ``Full Professor" in Telecommunications Engineering. Her work has appeared in first tier IEEE journals and she has received different awards, including the elevation to the grade of IEEE Senior Member in 2016, most downloaded article, and most cited article awards, and outstanding young faculty/researcher fellowships for conducting research abroad. Currently, Angela Sara serves as \textit{Area Editor} for IEEE Communications Letters, and as \textit{Editor/Associate Editor} for the journals: IEEE Trans. on Communications, IEEE Trans. on Wireless Communications, IEEE Open Journal of Communications Society and IEEE Trans. on Quantum Engineering. She was a recipient of the 2017 Exemplary Editor Award of the IEEE Communications Letters. In 2016 she has been an appointed member of the IEEE ComSoc Young Professionals Standing Committee. From 2017 to 2018, she has been the Award Co-Chair of the N2Women Board. Since 2017, she has been an elected Treasurer of the IEEE Women in Engineering (WIE) Affinity Group of the IEEE Italy Section. In 2018, she has been appointed as Publicity Chair of the IEEE ComSoc Women in Communications Engineering (WICE) Standing Committee. And since 2020, she is the vice-chair of WICE. Her current research interests are mainly in Quantum Communications, Quantum Networks and Quantum Information Processing.\par

\newpage

\begin{wrapfigure}{l}{25mm} 
    \includegraphics[width=1in,height=1.25in,clip,keepaspectratio]{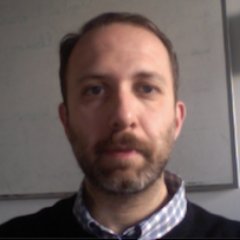}
  \end{wrapfigure}\par
  \textbf{Michele Amoretti} (S'01--M'06--SM'19) received his PhD in Information Technologies in 2006 from the University of Parma, Parma, Italy.
He is Associate Professor of Computer Engineering at the University of Parma.
In 2013, he was a Visiting Researcher at LIG Lab, in Grenoble, France.
He authored or co-authored over 100 research papers in refereed international journals, conference proceedings, and books. He serves as \textit{Associate Editor} for the journals: IEEE Trans. on Quantum Engineering and International Journal of Distributed Sensor Networks.
He is involved in the Quantum Information Science (QIS) research and teaching initiative at the University of Parma, where he leads the Quantum Software research unit.
He is the CINI Consortium delegate in the CEN-CENELEC Focus Group on Quantum Technologies. His current research interests are mainly in High Performance Computing, Quantum Computing, and the Internet of Things.\par

\vspace{0.2cm}

\begin{wrapfigure}{l}{25mm} 
    \includegraphics[width=1in,height=1.25in,clip,keepaspectratio]{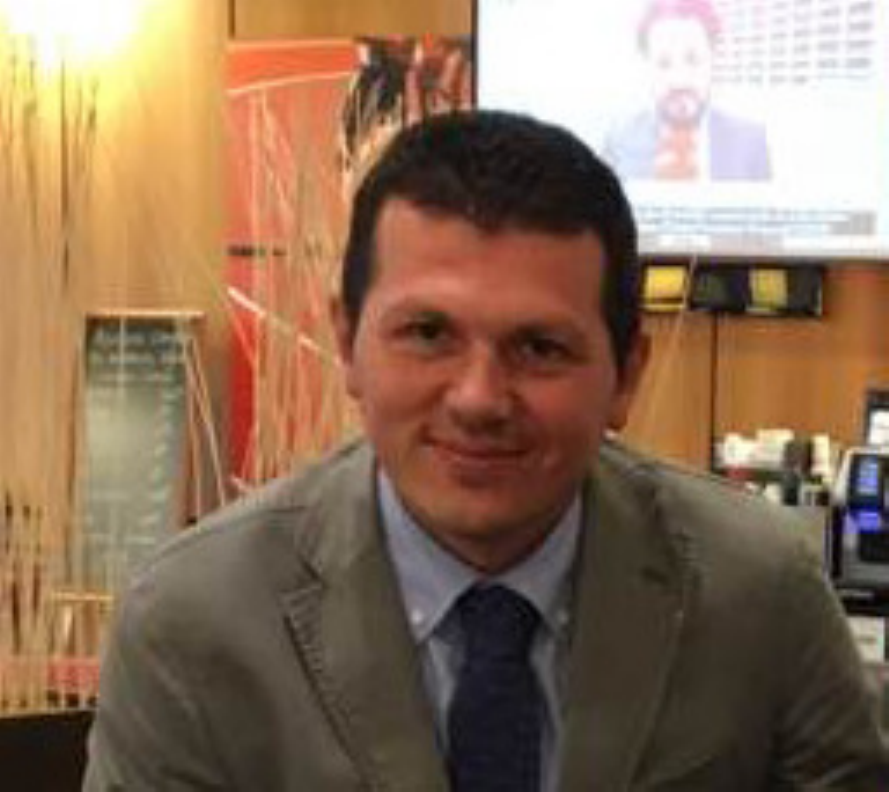}
  \end{wrapfigure}\par
  \textbf{Marcello Caleffi} (M'12, SM'16) received the M.S. degree (summa cum laude) in computer science engineering from the University of Lecce, Lecce, Italy, in 2005, and the Ph.D. degree in electronic and telecommunications engineering from the University of Naples Federico II, Naples, Italy, in 2009. Currently, he is with the DIETI Department, University of Naples Federico II, and with the National Laboratory of Multimedia Communications, National Inter-University Consortium for Telecommunications (CNIT). From 2010 to 2011, he was with the Broadband Wireless Networking Laboratory at Georgia Institute of Technology, Atlanta, as visiting researcher. In 2011, he was also with the NaNoNetworking Center in Catalunya (N3Cat) at the Universitat Politecnica de Catalunya (UPC), Barcelona, as visiting researcher. Since July 2018, he held the Italian national habilitation as \textit{Full Professor} in Telecommunications Engineering. His work appeared in several premier IEEE Transactions and Journals, and he received multiple awards, including \textit{best strategy} award, \textit{most downloaded article} awards and \textit{most cited article} awards. Currently, he serves as \textit{associate technical editor} for IEEE Communications Magazine and as \textit{editor} for IEEE Trans. on Quantum Engineering and IEEE Communications Letters. He served as Chair, TPC Chair, Session Chair, and TPC Member for several premier IEEE conferences. In 2016, he was elevated to IEEE Senior Member and in 2017 he has been appointed as Distinguished Lecturer from the \textit{IEEE Computer Society}. In December 2017, he has been elected Treasurer of the Joint \textit{IEEE VT/ComSoc Chapter Italy Section}. In December 2018, he has been appointed member of the IEEE \textit{New Initiatives Committee}.\par

\end{small}

\end{document}